\documentclass[twocolumn]{aa}
\usepackage{flushend}
\usepackage{graphicx}
\usepackage{txfonts}
\bibpunct{(}{)}{;}{a}{}{,} 
\usepackage{hyperref}
\usepackage{bm}
\usepackage{color}
\newcommand       \Ks           {{ K_{\rm S}}}

\begin{document}

\title{Distances to the Supernova Remnants in the Inner Disk
$^{}$\thanks{This paper is dedicated to the 60th anniversary of the Department of Astronomy, Beijing Normal University.}
}

\author{Shu Wang\inst{1}, Chengyu Zhang\inst{2}, Biwei Jiang\inst{2}, He Zhao\inst{2},
          Bingqiu Chen\inst{3}, Xiaodian Chen\inst{1}, Jian Gao\inst{2}, \and Jifeng Liu\inst{1}
          }
\institute{CAS Key Laboratory of Optical Astronomy,
                 National Astronomical Observatories,
                 Chinese Academy of Sciences, 
                 Beijing, 100101, China\\
         \and
                 Department of Astronomy, Beijing Normal University,
                 Beijing, 100875, China\\
        \email{bjiang@bnu.edu.cn}
         \and
                 South-Western Institute for Astronomy Research, Yunnan University,
                 Kunming, Yunnan 650091, China\\
             }

   \date{Received October 08, 2019; accepted May 14, 2020}


\abstract
{Distance measurements of supernova remnants (SNRs) are essential and important.
Accurate estimates of physical size, dust masses, and some other properties of SNRs depend critically on accurate distance measurements.
However, the determination of SNR distances is still a tough task. }
{Red clump stars (RCs) have a long history been used as standard candles.
In this work, we take RCs as tracers to determine the distances to a large group of SNRs in the inner disk.}
{We first select RC stars based on the near-infrared (IR) color-magnitude diagram (CMD). Then, the distance to and extinction of RC stars are calculated.
To extend the measurable range of distance, we combine near-IR photometric data from the 2MASS survey with the deeper UKIDSS and VVV surveys. With the help of the {\it Gaia} parallaxes, we also remove contaminants including dwarfs and giants.
Because an SN explosion compresses the surrounding interstellar medium, the SNR region would become denser and exhibit higher extinction than the surroundings. The distance of a SNR is then recognized by the position where the extinction and its gradient is higher than that of the ambient medium. 
}
{A total of 63 SNRs' distances in the Galactic inner disk are determined and divided into three Levels A, B, and C with decreasing reliability.
The distances to 43 SNRs are well determined with reliability A or B.
The diameters and dust masses of SNRs are estimated with the obtained distance and extinction.
}
{}


\titlerunning {Distances to SNRs in the Inner Disk}
\authorrunning {Wang et al. 2020}
\maketitle

\section{Introduction}

Distance is the essential parameter to further study of supernova remnants (SNRs). The diameter, brightness, age, and all the other properties of SNRs related depend sensitively on the distance. However, the determination of the SNR distance turns to be a tough task \citep{Ranasinghe2018MNRAS.477.2243R}.

The most common method of obtaining distances of SNRs is by analyzing the HI absorption spectra, which is based on the Galactic kinematics.
The rotation curve of the Galactic disk relates distance to the rotational velocity that may be determined from the radial velocity if the rotation is purely circular. \citet{Ilovaisky1972A&A....18..169I} first used this method to derive the kinematic distance for 20 SNRs. The ambiguity of this kinematical distance, i.e., each radial velocity corresponds to two distances equally spaced on either side of the tangent point, leads to large uncertainty. This method is improved in combination with the CO lines and applied to about 10 SNRs by \citet{Leahy2008A&A...480L..25L, Leahy2008AJ....135..167L, Leahy2010ASPC..438..365L, Tian2013ApJ...769L..17T}. Due to observational constraints, constructing reliable HI absorption spectra is still difficult.

The empirical power-law relation between the radio brightness and the linear diameter is also used to determine the distance of SNRs once the radio brightness is measured. This method suffers the errors from the dispersion of the empirical relation that the power-law index varies from about 2 to 6 \citep{Case1998ApJ...504..761C, Guseinov2003A&AT...22..273G, Pavlovic2013ApJS..204....4P}, and the index error can be as large as 40\% \citep{Zhu2014IAUS..296..378Z}.

There are some other methods to determine the distances of SNRs. The distance of the associated source in the SNR can be taken as the distance of the SNR \citep{Green1984MNRAS.209..449G}. Proper motion, in combination with expansion velocity, can derive the distance in case the SNR is close enough \citep{Green1984MNRAS.209..449G}. The X-ray flux can be used as a distance tracer as well \citep{Kassim1994ApJ...427L..95K}. These methods are usually dedicated to some specific cases.

The extinction toward an SNR nebula has already been used to measure the distance of SNRs.
A SNR is usually a dense cloud with high dust density due to three sources: (1) the circumstellar dust from the SN progenitor, (2) the interstellar dust swept and compressed by the SN explosion which is the major source for the SNRs in the Sedov-Taylor phase, and (3) the ejecta of the SN explosion. 
Hence,  the extinction of a SNR would increase sharply when the light from background stars passes through it. Recently, \citet{Zhao2018ApJ...855...12Z} measured the distance to the Monoceros SNR by identifying the position of the extinction jump of stars along the sightline. This accuracy depends on that of stellar distance and maybe about 10\%. This method requires precise measurement of stellar parameters of a large number of stars, usually from spectroscopic observation.

Another way to identify the distance of SNRs is based on the distance indicators -- red clump stars (RCs).
RCs were proposed as a standard candle early in 1998 \citep{Paczynski1998ApJ...494L.219P}. RCs are giants burning core helium with a stable structure which leads to almost constant luminosity and color. 
The dispersion of the absolute magnitude in the $K_{\rm S}$ band is about 0.03 mag, and the dispersion of the near-infrared (IR) color index $J-\Ks$ is about 0.02 mag \citep{Lopez-Corredoira2002A&A...394..883L}.
RCs are widely adopted to trace interstellar extinction as they can be easily distinguished in the color-magnitude diagram (CMD) \citep{Indebetouw_2005, Nishiyama_2006, Gao2009ApJ...707...89G, Wang_2017, Wang_Chen2019}.
With the known extinction of a target, its distance can be derived by measuring the distribution of extinction along the distance towards the sightline.
This idea was adopted to determine the distance to the neutron star in 4U 1608-52 by \citet{Guver2010ApJ...712..964G}. 
Later, \citet{Shan_2018, Shan_2019} used this method to estimate distances of SNRs with known extinction in the first and fourth quadrants of the Galaxy.

In this work, we try to use RCs to determine the distance to a large sample of SNRs in the inner disk. Basically, a supernova explosion compresses the surrounding material, which produces an SNR denser than the ambient medium. Consequently, the extinction around the SNR would increase more sharply than that of the ambient medium. Thus, we can compare the change of interstellar extinction toward the sightline of the SNR and the surrounding area to recognize the distance with a more sharp increase as that of the SNR. Briefly, we select the RCs from the near-IR CMD for both the SNR region and the surrounding area. Then, we calculate the corresponding extinctions and distances of the RCs.
Finally, we identify the distance of the SNR based on the apparent extinction jump of the RCs along the SNR sightline and the comparison with the ambient area.
Considering the complex environment in the inner disk, molecular clouds may lead to an increase of extinction as well.
Besides, it is plausible that SNRs are located in molecular clouds with physical interactions.
Hence, we scrutinize the distance to each SNR and evaluate the reliability of distance. 
To extend the measurable range of distance, we collect data not only from 2MASS, but also from UKIDSS, and VVV surveys. Combined with the {\it Gaia} parallaxes, we remove dwarf stars and red giants to compose the purer RCs samples. As a whole, we measured the distances to a group of 63 SNRs in the inner disk.
The specific skeleton is as follows.
The near-IR photometric data and SNRs data are described in Section \ref{Data}.
In Section \ref{Method}, we describe our method in detail.
The derived extinction of and distance to SNRs are presented in Section \ref{Result},
together with the distances' classification based on the reliability. 
Then, we compare our derived distance with previous works in Section \ref{Result}.
We analyze the extinction--diameter relation and estimate the dust mass of SNRs in Section \ref{Discussion}.
The feasibility and applicability of our method are discussed in Section \ref{Discussion} as well.
We summarize our principal results in Section \ref{Conclusion}.

\section{Data and Sample}\label{Data}

We collect the near-IR photometric data from the 2MASS, UKIDSS and VVV surveys, with the photometric quality  better than 0.05 mag in all bands and all catalogs. The UKIDSS and VVV surveys are brand new and deeper near-IR surveys relative to 2MASS.

\subsection{Near-IR Photometric Data}

\subsubsection{Fundamental Data: 2MASS}

Two Micron All Sky Survey (2MASS) is a whole-sky survey in the near-IR $JHKs$ bands \citep{Cohen2003AJ....126.1090C} by using two 1.3 m telescopes. 
The 2MASS PSC catalog contains over 470 million stars with 10 $\sigma$ limiting magnitudes of 15.8, 15.1, and 14.3 mag in the $J$, $H$, and $\Ks$ bands, respectively \citep{Skrutskie_2006}. Since SNRs are generally located in the Galactic plane suffering a relatively large extinction, such limiting magnitudes block the observation of distant SNRs. A deeper near-IR photometric survey would help. In this work, we use the data from the UKIDSS or VVV surveys as supplements.

\subsubsection{Supplemental Data: UKIDSS and VVV}

The Infrared Deep Sky Survey (UKIDSS) is a near-IR survey in the $J$, $H$, $K$ bands by using the 3.8 m United Kingdom Infra-Red Telescope (UKIRT) in Mauna Kea, which began in 2005 and observed 7500 $\deg^{2}$ in the northern sky. It includes five separate projects, and the data we use is from the Galactic Plane Survey (GPS), which covers approximately 1868 $\deg^{2}$ \citep{Lawrence2007MNRAS.379.1599L, Lucas2008MNRAS.391..136L}. The median 5 $\sigma$ point source depth in the $K$ band is 18.07 mag in the DR1 database of the GPS \citep{Warren2007MNRAS.375..213W}. This is about 4 magnitudes deeper than 2MASS and means a great extension to the distance range of SNRs traceable, while the order of extension depends on the specific interstellar extinction. On the other hand, this deep survey has a problem of saturation for the relatively bright sources.
\citet{Lucas2008MNRAS.391..136L} quoted that conservatively the saturation limit is best to assume to be 12th mag at $K$ band, although the typical saturation limit is $\thicksim$ 0.5 mag brighter.
For individual sightlines, we slightly adjust the saturation limit.

The VISTA Variables in The Via Lactea (VVV) is another near-IR survey by using the 4 m VISTA at ESO. This survey began in 2010, having completed 1929 hours observation, covering 520 $\deg^{2}$ in the Milky Way bulge and adjacent Milky Way plane. The VVV surveyed area is apparently in the southern sky due to the location of the telescope, which is complementary in space to the UKIDSS/GPS survey. The depth and saturation limits are comparable to the UKIDSS/GPS survey.  For the $J$ and $\Ks$ bands to be used \citep{Minniti2010NewA...15..433M},  the $\Ks$ limiting magnitude in the majority of regions is about 18.0 mag (5\,$\sigma$), while in the innermost field is about 16.5 mag \citep{Saito2012A&A...537A.107S}.
The $\Ks$ saturation magnitude for data with 10 s integrations is near 13 mag \citep{Soto2013A&A...552A.101S}, while for data with 4 s integrations is near 12 mag.
In this work, we adopt $\Ks \sim$ 12 mag as the limiting magnitude to avoid saturated VVV stars. The saturated stars in the VVV catalogues will be complemented with the 2MASS observations.

\subsubsection{Combination of the Fundamental and Supplemental Catalogs}

\begin{figure}[ht]
\centering
\includegraphics[width=\hsize]{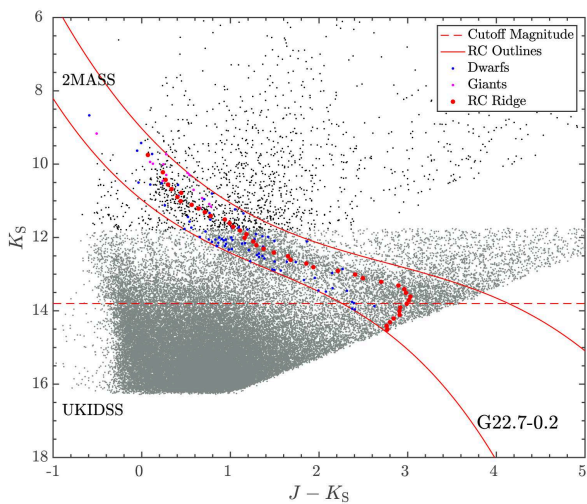}
\caption{CMD of the SNR G22.7-0.2. Black points denote the 2MASS data, and grey points denote the UKIDSS data. The borderline is at $K_{\rm S}$=11.8 mag.
Blue points mark dwarfs, and magenta points mark giants which are picked by comparing with the {\it Gaia} parallaxes (Section 3.2).
Red cubic curves roughly outline the range of RCs. The points locating between two red curves are thought to be RCs. The RC ridge is described by red points which is derived by the normal parameter estimation in this case. Red dashed line represents the cutoff magnitude which is 13.8 mag here.}
\label{FigCMD}
\end{figure}

In order to cover a wide range of distance, we combined the 2MASS data with the UKIDSS or VVV data. The 2MASS data will be used when the stars are brighter than the saturation limits of the other two deep surveys.
As the photometric systems are different in these surveys, we need to calibrate the photometric data into one standard.
Here, we convert the UKIDSS and VVV magnitudes to the 2MASS magnitudes using the transformation equations of \citet{Hodgkin2009MNRAS.394..675H} and \citet{Soto2013A&A...552A.101S} respectively.
Figure~\ref{FigCMD} is an example of the combined data for the SNR G22.7-0.2. It can be seen that the 2MASS data is smoothly adjoined with the UKIDSS data.

The complete set of CMDs for all the SNRs is appended in the Appendix as Figures \ref{FigUKICMD-1}, \ref{FigUKICMD-2}, \ref{FigVVVCMD-1}, \ref{FigVVVCMD-2}.

\subsection{SNRs Sample}

The sample of SNRs is from \citet{Green2014BASI...42...47G,Green2017yCat.7278....0G, Green2019JApA...40...36G} which compiled the parameters of up to about 300 SNRs. From the catalog, we select 159 SNRs and 124 SNRs that were observed  in the UKIDSS and VVV survey, respectively.

SNRs are extended objects and usually have an irregular appearance. To measure the extinction and distance of a SNR, the selected tracers, RCs, must be along the sightline of the SNR. Hence, the specific area of the SNR needs to be defined. On the one hand, a large area means more stars which can help to identify the RC branch in the CMD. On the other hand, a too large area may smooth out the feature of the SNR or contain additional feature beyond the SNR.
To validate our method to derive the distance of SNRs, only those with a large angular diameter, specifically $\geq 20'$, are selected. For the size, the angular radius defines the area for circular SNR. If a SNR is elliptical, still a circular area is defined with the semi-minor axis as the radius to minimize the contamination of other adjacent high extinction objects such as molecular clouds or star-forming regions.
With this size criterion, 35 SNRs with the UKIDSS data and 34 SNRs with the VVV data are kept, 6 of which are observed by both UKIDSS and VVV. Thus, our final sample contains 63 SNRs.

\section{Method} \label{Method}
\subsection{RC Candidates}

We use the red clump stars as the extinction and distance tracers, which are selected based on the $J-\Ks$ vs. $\Ks$ CMD. For each SNR area, the CMD consists of the stars from both the 2MASS catalog and UKIDSS or VVV catalog. Figure \ref{FigCMD} is an example of the CMD for G22.7-0.2. As described in the previous section, the bright stars are from the 2MASS catalog, and the faint stars are from the UKIDSS or VVV catalog, while the borderline depends on the sightline and is usually around $K_{\rm S}$=11.5-13 mag because of the saturation magnitude changes with the interstellar environment. In the case of G22.7-0.2 shown in Figure \ref{FigCMD}, the borderline is at $K_{\rm S}$=11.8 mag.

As RCs have an almost constant intrinsic color index and absolute magnitude in near-infrared, they would appear as a clump in the CMD without being affected by distance or extinction. In the CMD, the distance darkens the stars (shift vertically in CMD), while the extinction darkens and reddens the stars (shift diagonally in CMD). Hence, RCs appear like a branch from upper-left to lower-right instead of a clump in the CMD, which can be recognized by eyes.
However, in practice, the borders of the RC branch are not sharp. In the literature, people tried hard to find the exact position of the RCs in the CMD.
\citet{Guver2010ApJ...712..964G} outlined the borderlines of RCs according to the updated "SKY" model by \citet{Wainscoat1992ApJS...83..111W}. Then, RCs are extracted within the outlines.
After binning them by the $K$-band magnitude, they found the RC ridge by making the horizontal cuts and fitting RCs via a Gaussian function.
\citet{Saito2012A&A...537A.107S} fitted the RC ridge via a power-law plus a Gaussian to account for the contamination from dwarf stars.
To outline the RC branch, \citet{Lopez-Corredoira2002A&A...394..883L} fixed the width of the RC branch, while \citet{Gao2009ApJ...707...89G} improved by relaxing the width to a free parameter.  \citet{Gao2009ApJ...707...89G} took a quadratic polynomial curve to outline the RCs and fitted them by a Gaussian function.
Due to the influence of extinction on both observed color and magnitude and errors in photometry, the RC branch has a width increasing with magnitude, which means  that the \citet{Gao2009ApJ...707...89G} method is more reasonable.

The method we adopted to select RC candidates is on the basis of \citet{Gao2009ApJ...707...89G}.
The rough outlines of the RC branch are delineated by two cubic polynomial functions which fit 7 points selected manually. For an accurate determination of the RC ridge of this branch, a faint-ward cutoff is applied to the $\Ks$-band magnitude.
As shown in Figure~\ref{FigCMD}, the cutoff is at $\Ks = 13.8$ mag because the RC branch lacks enough red stars to form a complete branch on the red side when $K_{\rm S}>$ 13.8 mag. The RC ridge of the branch is searched within the outlines above the cutoff magnitude. Because the stellar density decreases with the $K_{\rm S}$ brightness, we dichotomize the bin in the $\Ks$ magnitude. At $K_{\rm S}>$ 10 mag, every 25 stars form a bin. At $K_{\rm S}<$10 mag, every 0.1 mag in $\Ks$-band is a bin.
For each bin, the RC ridge is obtained by the kernel density estimation (KDE) or the normal parameter estimation (Normal).
The Normal estimation is adopted when the result by KDE has excessive fluctuation.
The corresponding $J-\Ks$ and $\Ks$ values of the RC ridge are the average values of stars in each bin.
The RC ridge is then used to measure the distance and extinction of the SNR.

\subsection{Removal of Contaminants}

The selected RC sample within the two borderlines of the CMD contains some contaminants, mainly giants and dwarfs.
The near-IR intrinsic colors of red giants are similar to those of RCs. With the influence of interstellar extinction, red giants will mix with RCs in the CMD.
The low-mass dwarfs have red colors comparable to RCs.
\citet{Jian2017AJ....153....5J} found that the intrinsic color $(J-\Ks)_0$ of dwarf stars becomes redder than 0.62 mag at $T_{\rm eff} < 4500$ K. It means K-type and M-type dwarfs have intrinsic colors similar to or redder than RCs. Moreover, interstellar extinction also brings about redder color. As a result, the discrimination of giants and dwarfs from RCs is impossible simply from the CMD.

\begin{figure}[ht]
\centering
\includegraphics[width=\hsize]{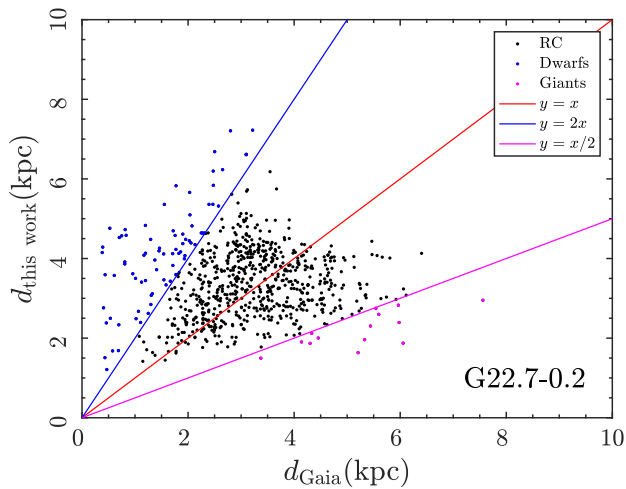}
\caption{The comparison of distance based on the RC assumption with the  {\it Gaia} distance for stars within the RC strip for G22.7-0.2 (see Figure \ref{FigCMD}). We use $y=2x$ (blue) and $y=x/2$ (magenta) lines to remove dwarfs and giants, respectively. Black points denote RCs, while blue points and magenta points denote dwarfs and giants, respectively. These contaminants are also marked in Figure \ref{FigCMD}.}
\label{FigDis-Dis}
\end{figure}

The {\it Gaia} mission \citep{Gaia2016A&A...595A...1G} is designed explicitly for astrometry. The recently released DR2 contains parallaxes for over 1 billion sources with high precision.
With the help of the {\it Gaia} parallaxes, some dwarf and giant stars can be excluded. In comparison to RCs, dwarfs are much fainter, and red giants are brighter. So the distance would be over-estimated if a dwarf star was mistaken as a RC and under-estimated if a giant star was mistaken as a RC.
To remove contaminations, we first calculate the distance of each star in the selected sample under the assumption of being a RC in the way to be described in Section 3.3.
Then, we compare the derived RC distance with the distance converted from the {\it Gaia} parallaxes with corrections by \citet{Bailer-Jones2018AJ....156...58B}.

Figure \ref{FigDis-Dis} is an example of the result for the field G22.7-0.2. Some stars have the {\it Gaia} distance significantly smaller than the RCs distance, and they must be dwarf stars mistaken as RCs.
Considering the errors (no constraint was set on the error of the {\it Gaia} distance) in both distances, the stars above the $y=2x$ line (i.e., the calculated RC distance is larger than twice the {\it Gaia} distance) are regarded as dwarf stars and excluded from the sample. Similarly, the stars below the $y=x/2$ line (i.e., the {\it Gaia} distance is larger than twice the calculated RC distance) are regarded as giant stars and excluded\footnote{The selection of using $y=2x$ and $y=x/2$ lines to remove dwarfs and giants is suitable. The line with other slope is  tried, but the RC ridge is not influenced much since the RC branch is apparent in the CMDs.}.
In this G22.7-0.2 example, 77 stars are classified as dwarf stars (1.5\%), and 14  stars are classified into giant stars (0.28 \%), and they are removed from the RCs sample.
These identified dwarfs (blue dots) and giants (red giants) are  marked in the CMD, as shown in Figure \ref{FigCMD}.
The complete set of distance comparison diagrams can be found in the Appendix as Figures \ref{FigUKIDD-1}, \ref{FigUKIDD-2}, \ref{FigVVVDD-1}, \ref{FigVVVDD-2}.
In some cases, the proportion of dwarfs is slightly high, due to the broad outlines of the RC branch in the lower part of CMD.
Even though this process to remove contaminants is imperfect, our procedure to select RC stars enables us to fit a locus that closely follows the RC branch in the CMD.

\subsection{Extinction and Distance of RCs}

As a standard candle, a RC star's distance can be determined by definition, $5\log d = m_\Ks - M_\Ks - A_\Ks + 5$. The reason to choose the $\Ks$ band is that it has higher photometric precision and a smaller scatter in the absolute magnitude comparing to the $J$ and $H$ bands.
The absolute magnitude $M_\Ks$ has a constant value reported in various works with small dispersion. Here we adopt $M_\Ks=-1.61$ mag \citep{Alves_2000, Ruiz-Dern_2018}.
The distance $d$ can then be calculated if the $\Ks$-band extinction $A_\Ks$ is known.
The extinction $A_\Ks$ can be derived from the color excess $E(J-\Ks)$ multiplied by a coefficient, $A_\Ks=c_\mathrm{e}\times E(J-\Ks)$. The coefficient $c_\mathrm{e}$ comes from the near-IR extinction law that defines the ratio of $A_\Ks/E(J-\Ks)$.

As discussed in \citet{Wang2014ApJ...788L..12W} and \citet{Wang_Chen2019}, the average near-IR extinction law is universal in most cases and follows a power law $A_{\lambda} \propto \lambda^{-2.07}$.
However, some literature reported that the extinction curve towards the inner Milky Way, including the Galactic center and the nuclear bulge, is variable and non-standard\citep[e.g.,][]{Goaling2009,Nataf2016MNRAS.456.2692N,2017ApJ...849L..13A}. 
A large power-law index is reported by using the deep near-IR observations, such as $\alpha=2.47$ \citep{2017ApJ...849L..13A} or $\alpha=2.64$ \citep{Goaling2009}. Recently, \citet{ChenXD2018} used Cepheids to investigate the near-IR extinction along the sightlines to the Galactic center region and found that the variation in near-IR extinction law is small for sightlines to the Galactic center and nearby regions even if the absolute extinction varies a lot.
The corresponding power-law index is $\alpha=2.05$ with the VVV effective wavelengths as benchmarks.
This value is consistent with $\alpha=2.11$ reported by \citet{Fritz2011}, which is determined based on hydrogen emission lines in the Galactic center.
Besides, \citet{Wang_Chen2019} discussed the extinction uncertainties brought from the measurement method in details (see Section 4 of their work) and concluded that the reported various near-IR extinction in previous works could be explained by several factors including the photometric quality, the average reddening amount of high-extinction sources, the sample number, and the ratio of low- to high-extinction sources. 
For high-extinction regions, such as the Galactic center and molecular clouds, the curvature of color excess ratios becomes obvious, which also affects the measurement of the near-IR extinction law.
As the SNRs in our sample have a size larger than $20\arcmin$, the average IR extinction law is acceptable.
In this work, we adopted the coefficient, $c_\mathrm{e}=0.473$, of \citet{Wang_Chen2019}, which is derived based on RCs with the {\it Gaia} data.
The color excess $E(J-Ks)$ is obtained by subtracting the intrinsic color index  $(J-\Ks)_0=0.7$ mag \citep{Grocholski2002AJ....123.1603G}. The distances of each RC star and the RC ridge are then determined from
\begin{equation}\label{equ1}
5\log d = m_\Ks-(-1.61)-0.473\times(J-\Ks-0.7)+5 ~~~.
\end{equation}

For each RC star, if assuming the near-IR extinction law uncertainty is $\sim$ 15\%
\footnote{The 15\% uncertainty in the near-IR extinction law includes measurement errors of methods and the reported variations in the literature.}
at $E(J-\Ks)=2$ mag, the propagated distance uncertainty is $\sim$ 7\%.
To reduce uncertainties, we take the RC ridge as a typical value for each bin to estimate the extinction and distance of SNRs.

\subsection{Distances of SNRs}

\begin{figure}
\centering
\includegraphics[width=\hsize]{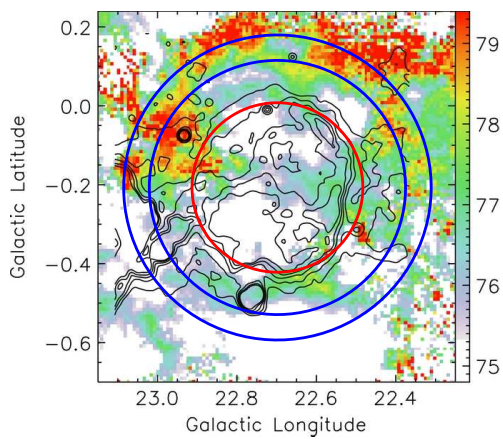}
\caption{The definition of the SNR area (the red circle) and the surrounding area (the blue annulus) for SNR G22.7-0.2 as an example.
The background image is the $^{13}$CO ($J$=1-0) emission map overlaid with the 1.4GHz radio continuum emission contours from \citet{Su2014ApJ...796..122S}.}
\label{Fig-glgb}
\end{figure}

\begin{figure}
\centering
\includegraphics[width=\hsize]{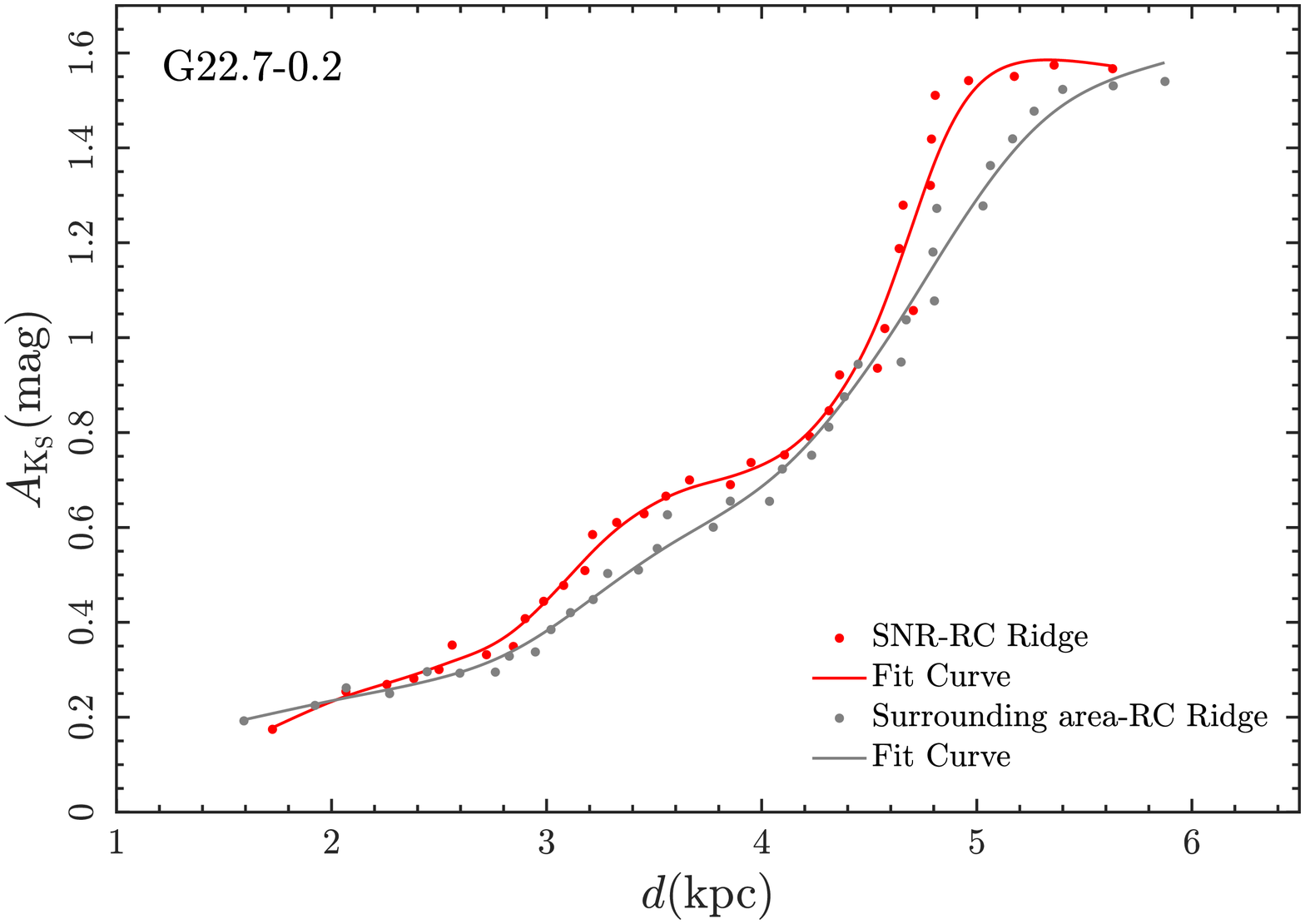}
\includegraphics[width=\hsize]{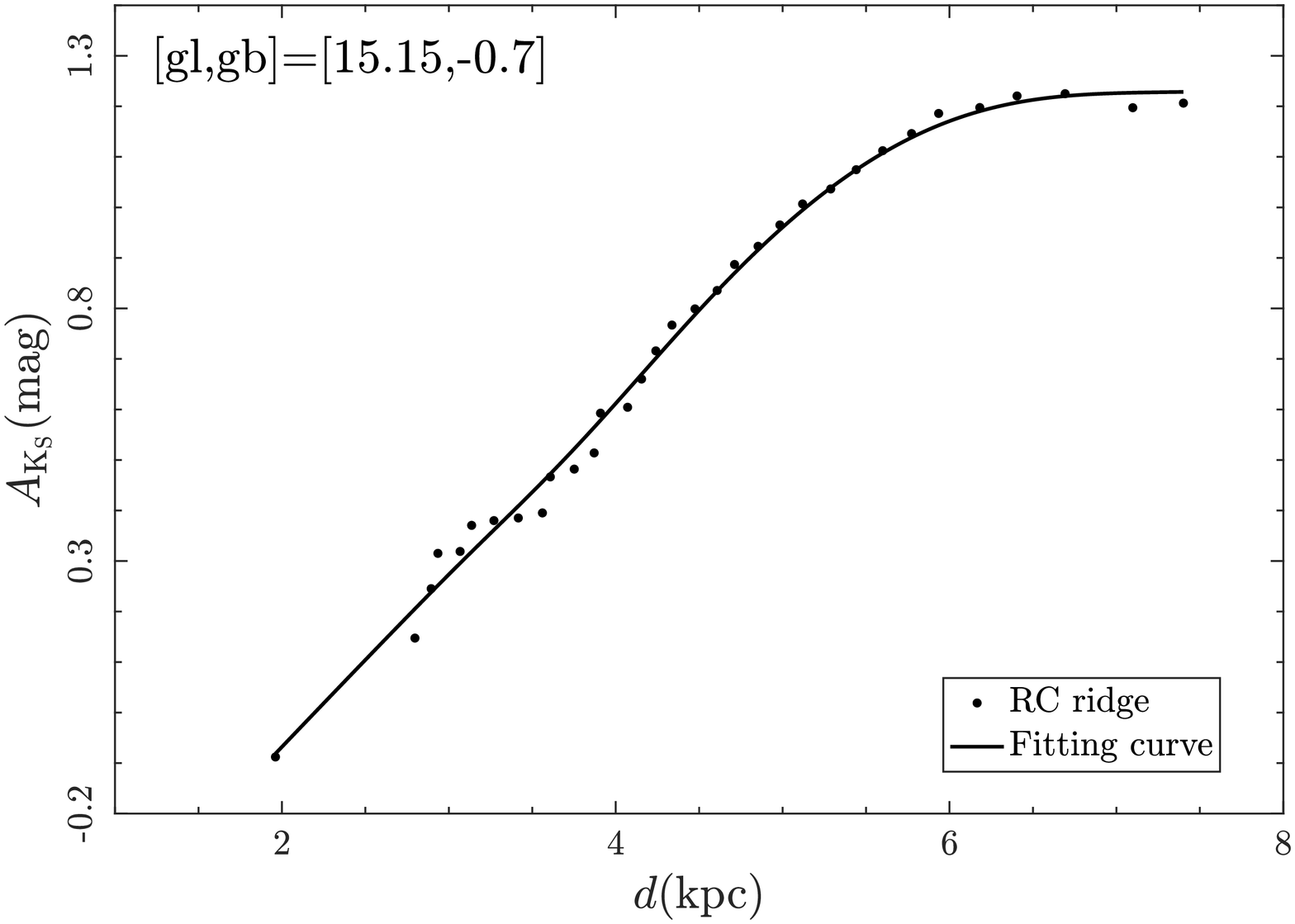}
\caption{(a) (Top): The distance--extinction diagram of SNR G22.7-0.2.
The red and grey dots denote the RC ridge of the SNR region and the surrounding region, respectively. The lines are the spline interpolation function curves.
There is a noticeable sharp jump at about 4.74 kpc, and a small jump at about 3.11 kpc.
(b) (Bottom): The distance--extinction diagram of the sightline of molecular cloud G15.15-0.7.
}
\label{FigD-A}
\end{figure}
\begin{figure}
\centering
\includegraphics[width=\hsize]{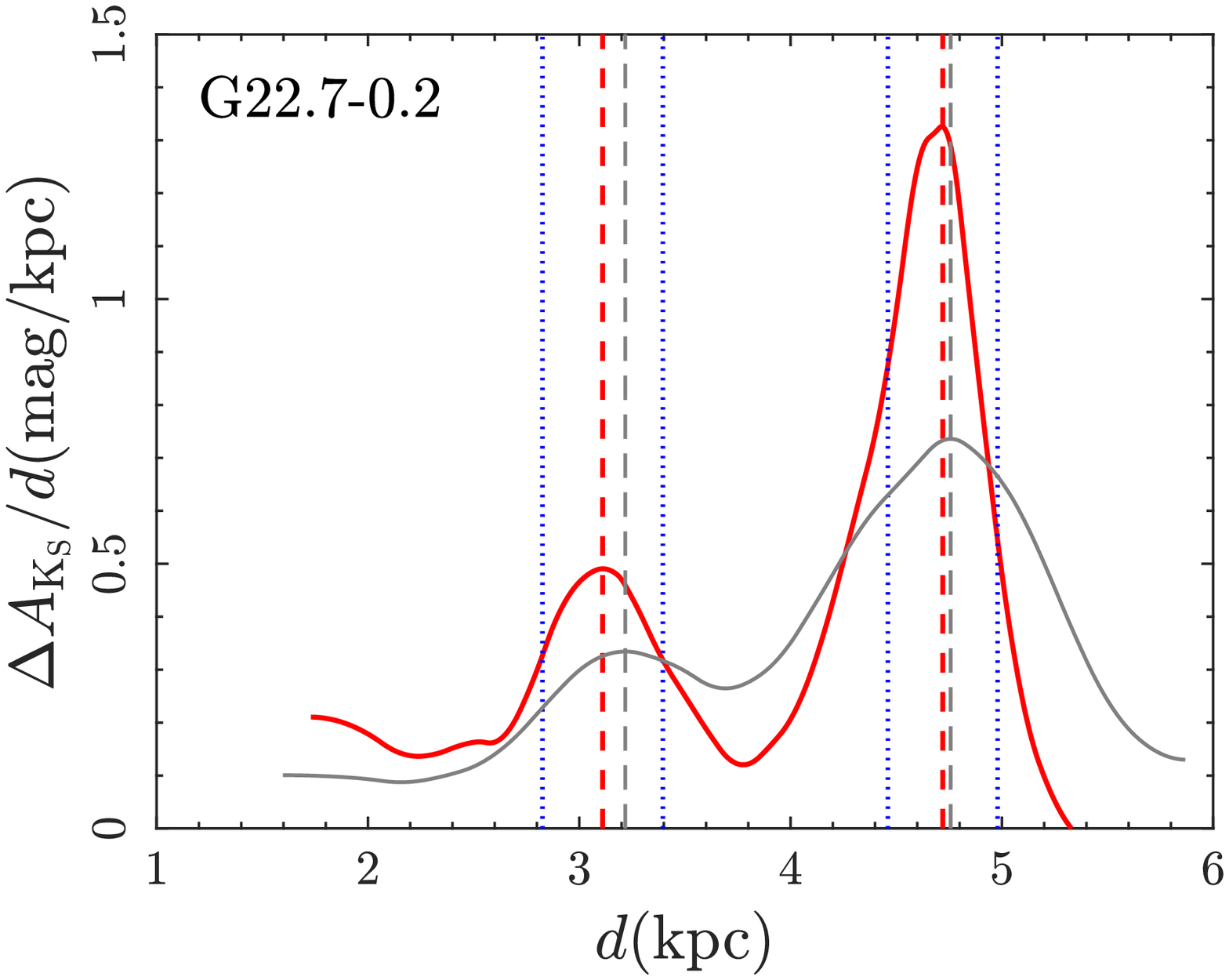}
\includegraphics[width=\hsize]{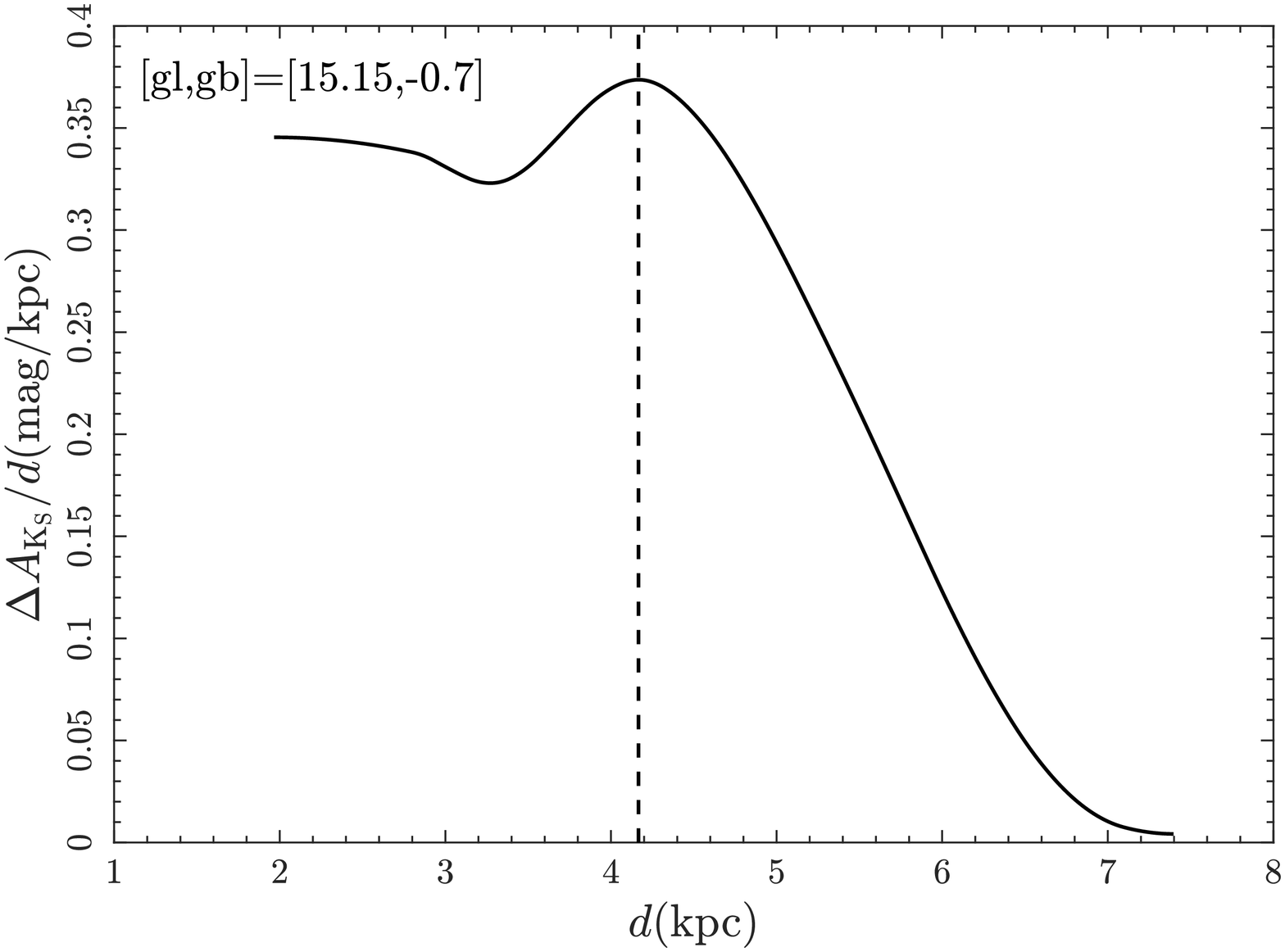}
\caption{(a) (Top):The change of differential extinction per distance $\Delta A_\Ks/d$ with distance for SNR G22.7-0.2 (red line) and the surrounding area (grey lines).
For the SNR sightline, there are two significant gradients marked as red dash lines:
the larger one at 4.74 kpc with $\Delta A_\Ks/d=1.33$ mag/kpc,
and the smaller one at 3.11 kpc with $\Delta A_\Ks/d=0.49$ mag/kpc.
The blue dotted lines indicate the locations of half maximum half width of the profile, which represent the distance errors. For the surrounding area, the extinction gradients are weaker and marked as the grey dash lines. (b) (Bottom): The distribution of differential extinction per distance for the sightline of molecular cloud G15.15-0.7.
}
\label{FigD-dA}
\end{figure}
The distance of a SNR is determined in two steps. As we mentioned in the Introduction section, the supernova explosion compresses the surrounding interstellar material to a high density so that the interstellar extinction would increase sharply at the position of the SNR. Then, the first step searches for the position(s) where a sharp extinction gradient occurs, which is the method used in \citet{Zhao2018ApJ...855...12Z, Zhao2020}. Considering that the SNRs are located in the Galactic plane full of molecular clouds, the second step tries to identify whether the distance is of the SNR itself or just of a molecular cloud in the sightline. The details follow.  

Firstly the change of $A_\Ks$  with the distance $d$ is studied for the RC ridge toward each SNR sightline. Taking G22.7-0.2 as an example, Figure \ref{FigD-A} shows the change of $A_\Ks$ with $d$. The general increasing trend of extinction with distance is significant. From this diagram, the position of the sharp extinction gradient can be recognized.
Though this position is usually visible, we use a smooth spline interpolation function to determine this position quantitatively, which is the red line in Figure \ref{FigD-A}, i.e. the spline interpolation curve of the red dots\footnote{In a few cases, there exist significant outliers that are removed in determining the spline interpolation curves.}.
The accurate positions of the two jumps are then calculated by differentiating the spline interpolation function. The result for G22.7-0.2 is displayed in Figure~\ref{FigD-dA},  i.e. the differential extinction per distance $\Delta A_\Ks/d$ with distance, where the two significant gradients are marked as red dash lines.
The primary gradient occurs at 4.74 kpc with a value of 1.33 mag/kpc, significantly larger than the average extinction rate around 0.05 mag/kpc in the diffuse ISM\footnote{In the solar neighborhood, the average extinction rate in the $V$ band is 0.7--1.0 mag/kpc \citep{1969ApJ...157..611G, 1980AJ.....85...17M}.
\citet{Wang_2017} estimated the average rate to be 0.37 mag/kpc towards a very diffuse region in the Galactic plane.
\citet{Wang_Chen2019} updated the average extinction law with improved precision, from which we adopt $A_\Ks/A_{\rm V}=0.078\pm 0.004$ and the corresponding average rate of extinction in the $\Ks$-band is 0.029--0.078 mag/kpc.}.
The secondary gradient occurs at 3.11 kpc with a value of 0.49 mag/kpc, which is also larger than the average value of diffuse clouds.
The blue dotted lines indicate the locations of half maximum half-width of the profile, which represent the distance errors. Consequently, two distance candidates are found for G22.7-0.2 by the sharp gradient that is significantly larger than the average. 

The second step tries to identify whether the distance candidate is of the SNR or just of a molecular cloud in the sightline. This is done by comparing the behavior of $A_\Ks$ and  $\Delta A_\Ks/d$. For this purpose, an area surrounding the SNR is selected for comparison.
The surrounding area is an annulus, where the inner radius of the annulus is 1.5 times the radius of the SNR to avoid the SNR's influence, and the area of the annulus is equal to that of the SNR so that there would be a comparable number of stellar tracers in the surrounding area to the SNR region. It is expected that the SNR would produce more extinction and a sharper gradient due to its higher dust density than the surrounding area. Figure~\ref{Fig-glgb} shows the case of G22.7-0.2, where the red circle marks the SNR G22.7-0.2, and the blue annulus is the surrounding area.
The background image is the weighted intensity mean velocity map of the $^{13}$CO ($J$=1-0) emission overlaid with the 1.4GHz radio continuum emission contours from \citet{Su2014ApJ...796..122S}. In Figure~\ref{FigD-A} and \ref{FigD-dA}, the grey lines show the change of $A_\Ks$ and  $\Delta A_\Ks/d$ with $d$ for the surrounding area. It can be seen that the SNR G22.7-0.2 presents both higher $A_\Ks$ and  $\Delta A_\Ks/d$ at the candidate distances as expected. In such a case, the candidate distance is regarded as the distance of the SNR. There are some cases where the SNR region shows no excess or even smaller extinction or gradient in comparison with the surrounding area, then the candidate distance may not be associated with the SNR and is not accepted.

It may be argued that the high extinction and gradient can be caused by the molecular cloud in the sightline. To illustrate that a molecular cloud produces very likely different scenario, we chose one intense region located at $l=15^\circ.15$, $b=-0^\circ.7$ with radius=0.25$^\circ$ (named G15.15-0.7) from the Planck 100GHz map that is representative of the molecular CO \citep{2018arXiv180706205P}. Its change of the extinction and the gradient with distance are shown in Figure~\ref{FigD-A} and Figure~\ref{FigD-dA}. It can be seen that the change is much more smooth and correspondingly the gradient is smaller though there is still a peak around 4.2 kpc. The situation is very similar to that of the surrounding area of G22.7-0.2. 
This shows that typical molecular clouds do not produce a large increase of extinction in a narrow distance range like an SNR, but the existence of two peaks in Figure~\ref{FigD-dA} shows that this can happen in rare cases, see also Figure~\ref{FigUKIdAk-1}.
This can be understood that a normal molecular cloud has a lower density than a SNR. There may be dense cores within a molecular cloud, but a dense core would be very obscured, and few RCs could be observable and presented in the analysis. In addition, a dense core is rather small, and its effect may be smoothed out in size like a SNR which is larger than 20$\arcmin$ in our sample. Therefore, in principle, it is reasonable that we accept the distance at the position of a gradient larger than the surrounding area toward the sightline of the SNR, but the study of all types of molecular clouds is beyond the scope of this paper.

A further step is taken to confirm the identified distance and to distinguish which distance is associated with the SNR when more than one candidate is found, i.e. to check if the SNR was found to interact with any molecular cloud from previous studies, similar to the method of \citet{Zhao2020}. If the SNR is interacting with the molecular cloud, the distances of the two objects are the same. Then, the distance of the SNR is determined. In the case of G22.7-0.2, \citet{2010ApJ...712.1147J} suggested it is in physical contact with the environmental molecular cloud. \citet{Su2014ApJ...796..122S} identified a molecular cloud complex GMC G23.0-0.4 that covers SNR G22.7-0.2 (see their Figure 1) based on their CO observations and found the convincing kinematic evidence that supports the interaction between the SNR and the 77 km s$^{-1}$ molecular clouds.
According to \citet{2010ApJ...712.1147J}, the kinematic distance of G22.7-0.2 is $4.4\pm0.4$ kpc, which coincides with the position of the primary extinction gradient at 4.74 kpc in our calculation while disagrees with the position of the secondary gradient at 3.11 kpc which may be caused by another foreground dense cloud. Therefore, the distance of G22.7-0.2 is determined to be 4.74 kpc, which confirms previous identification of the interaction and is more accurate. The details for the individual objects can be found in Section \ref{Result}.

The complete set of distance-extinction diagrams for all the SNRs is shown in the Appendix as Figures \ref{FigUKIAk-1}, \ref{FigUKIAk-2}, \ref{FigVVVAk-1}, \ref{FigVVVAk-2}.
The distributions of differential extinction per distance with distance for all the SNRs are displayed in the Appendix as
Figures \ref{FigUKIdAk-1}, \ref{FigUKIdAk-2}, \ref{FigVVVdAk-1}, and \ref{FigVVVdAk-2}.
In case that a SNR has two distance choices, such as G22.7-0.2, additional information, such as the known distances from previous works and radio observations with velocity, is used to judge which distance is that of the foreground/background molecular cloud and the SNR.

\section{Result}\label{Result}

The distance and extinction of 63 SNRs based on the 2MASS and UKIDSS or VVV data are derived.
We tabulated these results in Table \ref{UKIDSStable} and Table \ref{VVVtable}, including name, the other name for famous SNRs,  position, radius, the adopted fitting method to derive the RC ridge,  distance, and extinction.
We named the distance determined in this work as the extinction distance $d_{\rm ext}$.
If the sightline of a SNR exhibits two distance gradients, the larger one is assigned to the primary distance $(d_{\rm ext})_{\rm P}$, and the smaller one is assigned to the secondary distance $(d_{\rm ext})_{\rm S}$.
The finally suggested extinction distances to SNRs are in boldface.
Generally, the errors of distances in Table \ref{VVVtable} (based on VVV data) are smaller than those in Table \ref{UKIDSStable} (based on UKIDSS data).
Among 63 SNRs, 6 of them have both UKIDSS and VVV data, and  their distances have been determined independently by the same method with the different data samples.
They are:
\begin{itemize}
\item G5.4-1.2 with $d_{\rm ext}=3.89\pm0.91$ kpc (UKIDSS), $d_{\rm ext}=3.89\pm0.37$ kpc (VVV),
\item G6.1+1.2 with $d_{\rm ext}=3.27\pm0.73$ kpc (UKIDSS), $d_{\rm ext}=3.67\pm0.36$ kpc (VVV),
\item G6.4-0.1 with $d_{\rm ext}=3.55\pm0.90$ kpc (UKIDSS), $d_{\rm ext}=3.55\pm0.34$ kpc (VVV),
\item G8.9+0.4 with $d_{\rm ext}=3.54\pm0.62$ kpc (UKIDSS), $d_{\rm ext}=3.51\pm0.41$ kpc (VVV),
\item G359.0-0.9 with $d_{\rm ext}=3.49\pm0.36$ kpc (UKIDSS), $d_{\rm ext}=3.29\pm0.20$ kpc (VVV),
\item G359.1-0.5 with $d_{\rm ext}=3.29\pm0.47$ kpc (UKIDSS), $d_{\rm ext}=3.18\pm0.32$ kpc (VVV).
\end{itemize}
The agreement between these six sources is very good.
This high internal consistency underscores that our distance determination method is stable.

The extinction distance is identified based on the near-IR extinction jump of the RC ridge along the SNR sightline.
Due to the extinction effects, the distance can not be measured for the SNRs with either too low extinction or too high extinction.
The lower limit of measurable distance is about 2 kpc.
Depending on the amount of extinction in the sightline of each case, the upper limit of measurable distance is changeably ranging from 5 kpc to 16 kpc.
In this work, most SNRs with measurable extinction distances are located within 2--5 kpc.

Adopting this method to derive extinction distance may be uncertain in two situations.
The first case is that there is no sharp jump if the extinction increases slowly with the distance. For example, G59.8+1.2 (Figure~\ref{FigUKIdAk-2}) only shows a  small amplitude of variation in the extinction gradient, which neither presents any additional extinction or gradient in comparison with the surrounding area. This phenomenon may indicate that the SNR is very tenuous and almost completely dispersed into the interstellar medium, or becomes filamentary so that most of the stars experience no additional extinction to the surrounding interstellar medium.
The second case is that the relatively sharp increase of the extinction with distance spans a range significantly larger than the typical size of a SNR and the error of distance. G6.4-0.1 is one such case that the sharp increase starts from $\sim$\,2.4\,kpc and ends at $\sim$\,5.4\,kpc (Figure~\ref{FigUKIdAk-2}). The about 3\,kpc width cannot be caused by a single SNR, but possibly by some giant molecular clouds and/or adjacent SNRs and clouds, which looks like the profiles of the comparison molecular cloud G15.15-0.7 in Figure~\ref{FigD-dA}.
Indeed, G6.4-0.1 is known to have strong kinematic evidence for the SNR--molecular cloud interaction \citep{2010ApJ...712.1147J}. It is an old-age SNR with an estimated distance between 1.8 and 3.3 kpc \citep[e.g.,][]{Goudis1976Ap&SS..40...91G,Lozinskaya1981SvAL....7...17L,Velazquez2002AJ....124.2145V,Aharonian2008A&A...481..401A}. Both effects are reflected in the profile width of the extinction gradient that is taken as the uncertainty of the distance in Table~\ref{UKIDSStable} and~\ref{VVVtable}.

\subsection{Reliability Levels of Distances}

The reliability of the derived distances is divided into three levels. The most reliable result is required to satisfy the following criteria at the candidate distance: (1) the extinction gradient  $\Delta A_\Ks/d \geq 0.09$ mag/kpc, i.e. greater than the average gradient of the diffuse interstellar medium; (2) the extinction $A_\Ks$ of the SNR region is greater than the surrounding area, which means some excess extinction in comparison with the surrounding area; (3) the extinction gradient  $\Delta A_\Ks/d $ of the SNR region is greater than the surrounding area; (4) the width of the $\Delta A_\Ks/d $ vs. $d$ profile is less than 20\% of $d$, i.e. a kind of relative error, which means a relatively small size of the structure. In spite that some molecular clouds are small as well, this criterion excludes large molecular clouds. 
The sightline toward SNRs in the inner disk may contain several clouds around the position of spiral arms. In such cases, the identification of the distance to the SNR is scrutinized in Section~\ref{SNRs_distribution}.
If all the above four criteria are satisfied, the distance has the highest reliability, Level A. Consequently, 34 of the 63 SNRs are at this level.

The other 29 SNRs are further divided into two levels, Level B and C, depending on whether the SNR is found to be associated with any molecular cloud.
\citet{2010ApJ...712.1147J} presented a catalog of 64 Galactic SNRs known and suggested to be in physical contact with environmental molecular clouds.
\citet{Jeong2012} found an additional 6 SNRs having spatial correlations with molecular clouds.
Recently, \citet{2019ApJ...884..113S} also tabulated the SNRs with evidence of interaction with a molecular cloud.
\citet{2015MNRAS.454.2586F} presented 30 SNRs with H$_2$ emission features, and
the detailed work about these SNR interacting with molecular clouds is presented in \citet{2019AJ....157..123L}. By cross-checking, 23 SNRs in our sample are found to be in these catalogs (marked in Tables \ref{UKIDSStable} and  \ref{VVVtable}), which means they are associated with some molecular clouds. Twelve of them satisfy the above criteria and are already classified into Level A. For the other 11 SNRs, the derived extinction distance is examined to be consistent with the kinematic distance of the cloud or not. If yes, this distance is classified into Level B, otherwise, into Level C, which results in 9 SNRs in Level B and 2 SNRs in Level C. 
In such a way, the independently determined distances in Level B are highly reliable since the distance of the SNR is consistent with the associated molecular clouds. 
These SNRs may be old and filamentary, which cannot show up either in the total extinction or the extinction gradient in comparison with the surrounding area, so that they are not in Level A.  All the other 18 SNRs, which neither satisfy the criteria or found to be associated with any molecular cloud, are classified into Level C. The distances in Level C still have some chances to be real and can be taken as a reference since not all the SNRs are carefully checked to interact with a molecular cloud or not. 
The final classification of distances is listed in the column of "Reliability'' in Table \ref{UKIDSStable} and Table \ref{VVVtable}.
To summarize, our determined distances to 63 SNRs in the inner disk are classified into three groups, A, B, and C, including 34 highly reliable (A), 9 reliable (B) associated with molecular clouds, and 20 less reliable (C) which needs further investigation.
The reliable distances (A and B) are derived for more than two-thirds of 63 SNRs. The accurate distances to seven SNRs\footnote{G5.4-1.2, G308.8-0.1, G318.2+0.1, G318.9+0.4, G327.1-1.1, G329.7+0.4, and G341.2+0.9} are obtained for the first time. For those SNRs associated with molecular clouds, the derived distance confirms or distinguishes previous distances, and with higher accuracy.

\subsection{Comparison with Previous Works}\label{comp}

Currently, the known distances of SNRs mostly are the kinematic distances or the radio-surface-brightness distances.
The kinematic distance is determined by data from HI and CO line surveys.
The radio-surface-brightness distance is usually estimated from the radio-surface-brightness--diameter relation. 
In our sample, there are 12 SNRs with both the kinematic distances $d_{\rm kin}$ from \citet{Green2019JApA...40...36G} and the radio-surface-brightness distances $d_{\rm rad}$ from \citet{Pavlovic2013ApJS..204....4P}.
Only one source, G93.7-0.2, has consistent $d_{\rm kin}$ and $d_{\rm rad}$.
Besides, the $d_{\rm rad}$ value of G327.4+0.4 is close to the lower limit of the $d_{\rm kin}$ range.
The distances of three cases, G22.7-0.2, G23.3-0.3, and G353.6-0.7, are consistent within $\sim$\,1.5\,kpc.
The discrepancies of the other 7 SNRs are larger than 2.0\,kpc, and the most substantial difference even reaches 6.4\,kpc.
The recent works of \citet{2019ApJ...884..113S} and \citet{2019SerAJ.199...23S}, also tabulated the distances to a large group of SNRs. We compare these distances with the radio-surface-brightness distances as well.
In \citet{2019ApJ...884..113S}, the adopted distances to G321.9-0.3, G335.2+0.1, G351.7+0.8 are $6.5^{+3.5}_{-1.0}$ kpc, 1.8 kpc, $13.2\pm0.5$ kpc, respectively, whereas the $d_{\rm rad}$ distances from \citet{Pavlovic2013ApJS..204....4P} are 3.8 kpc, 4.2 kpc, 5.4 kpc.
The discrepancy is apparent, in particular as large as 7.8 kpc for G351.7+0.8.
SNR G351.7+0.8 is a shell-like SNR, whose distance is estimated to be $13.2\pm0.5$ kpc based on an associated HI emission by \citet{2007MNRAS.378.1283T}, and 5.4 kpc using the radio surface-brightness--diameter relationship by \citet{Pavlovic2013ApJS..204....4P}.
In \citet{2019SerAJ.199...23S}, the adopted distances to G15.1-1.6, G73.9+0.9, G359.1-0.5 are 2.2 kpc, 1.25 kpc, 8.5 kpc, respectively, also apparently different from their $d_{\rm rad}$ distances from \citet{Pavlovic2013ApJS..204....4P} being 4.2 kpc, 4.0 kpc, 4.0 kpc.
Moreover, the distances to some SNRs are still controversial, such as SNR G359.1-0.5.
As summarized in \citet{2020arXiv200307576S} and \citet{2020MNRAS.493.3947E}, there are suggestions that the SNR located at the Galactic Centre, 8.5 kpc, or the foreground of the Galactic center, $\sim4$ kpc.
In addition, even the kinematic distances in the literature are inconsistent with each other.
For example, in Green's SNRs catalog, the distance of G347.3-0.5 implied by the associated molecular clouds and X-ray observations is 1.3 kpc.
Meanwhile, \citet{1999ApJ...525..357S} argued that a distance as small as this appears to be very unlikely.
Based on several lines of evidence, including a distance estimation based on the X-ray derived column density, a quantitative calculation based on the rotation curve of the Galaxy, and a complete picture of the remnant, they suggested a distance of 6 kpc with at least $\pm1$ kpc uncertainty for G347.3-0.5.

\begin{figure}
\centering
\includegraphics[width=4.2in]{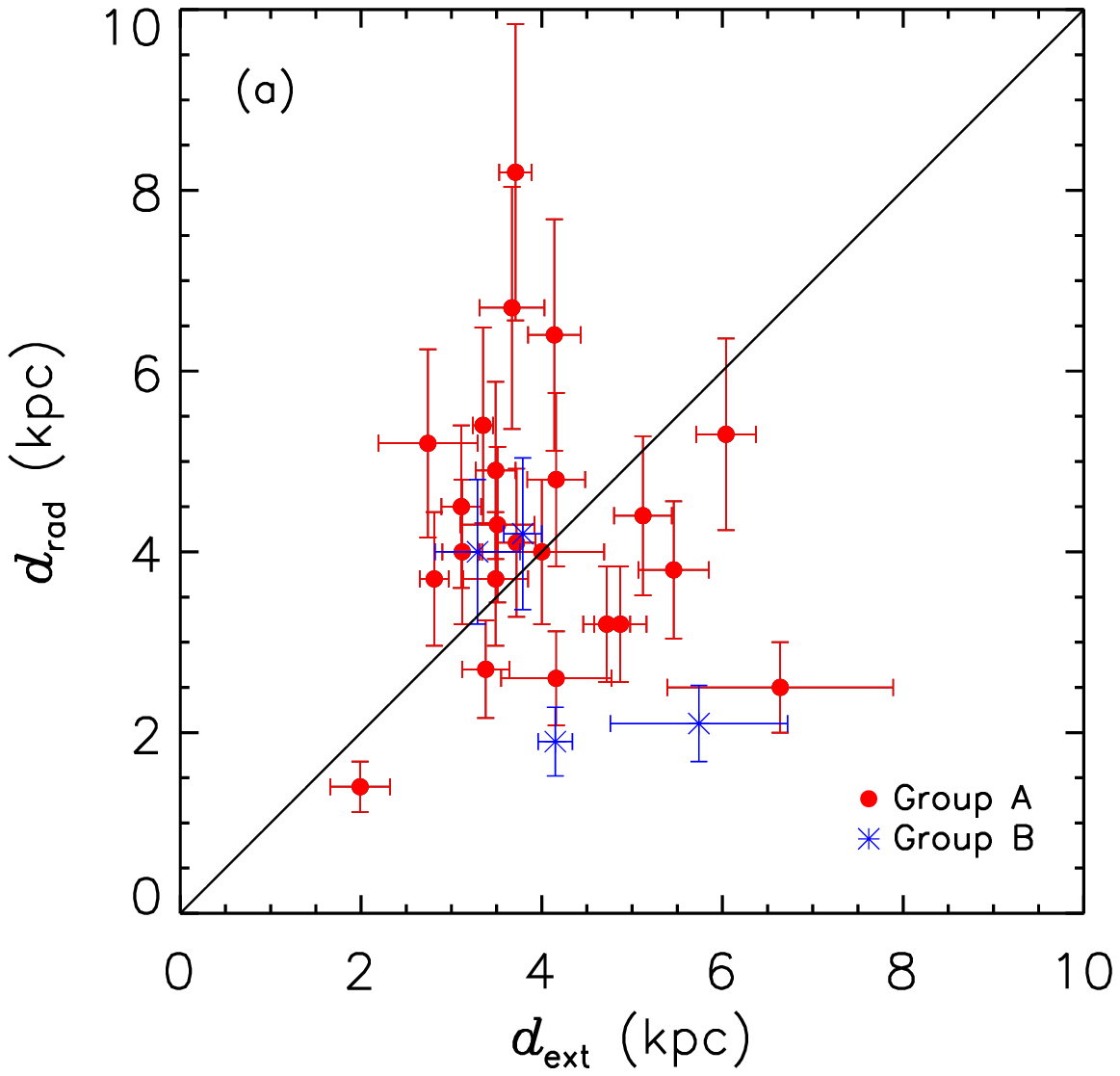}
\includegraphics[width=4.2in]{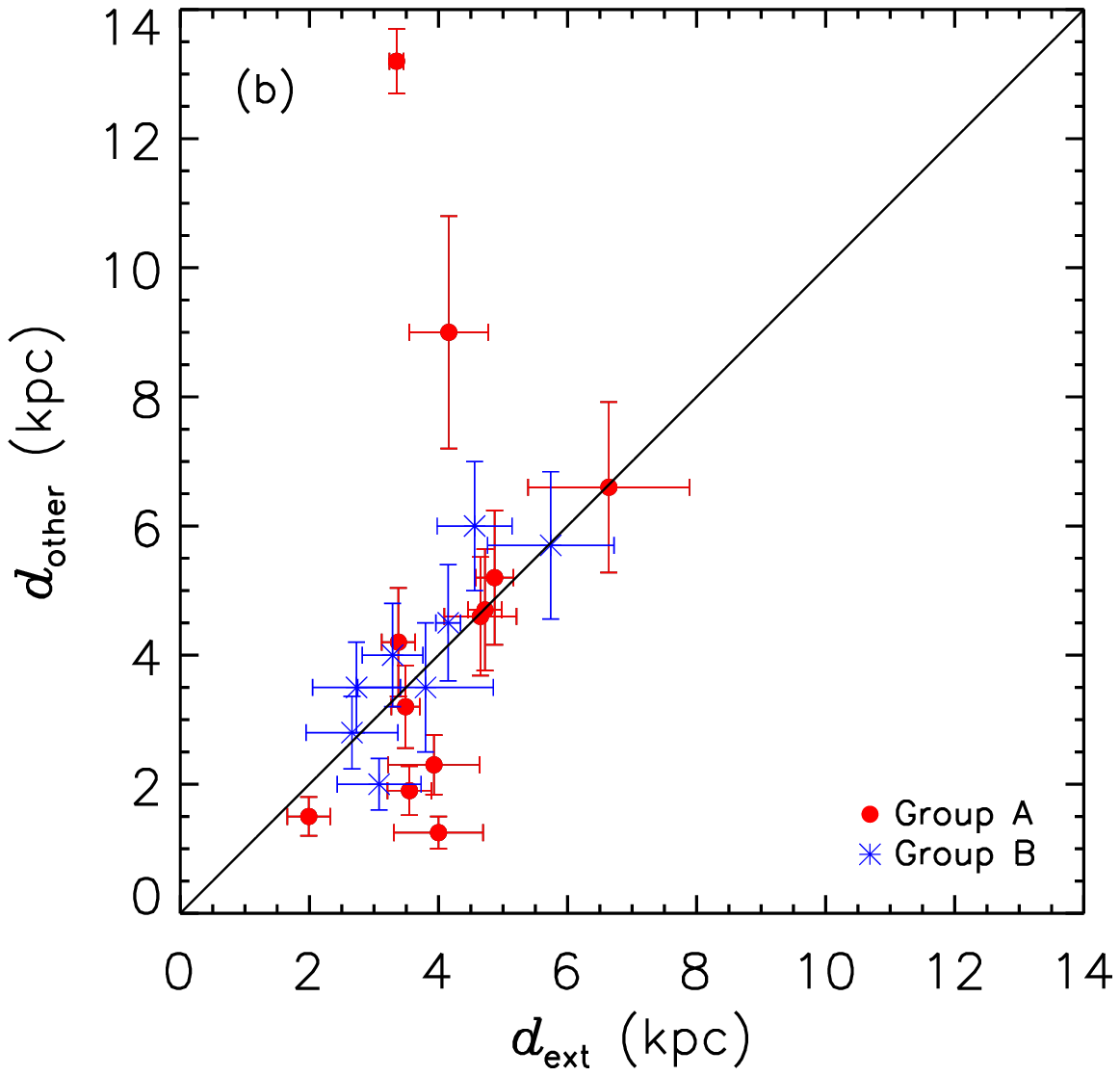}
\caption{Comparison of distances derived in this work (extinction distance $d_{\rm ext}$) with
(a) (Top) the radio-surface-brightness distances $d_{\rm rad}$ for 27 SNRs,
and (b) (Bottom) the distances obtained by other methods $d_{\rm other}$ for 20 SNRs.
The red dots and blue asterisks denote the SNRs with distance's reliability A and B, respectively.
The solid lines are the y=x loci.}
\label{comp_d}
\end{figure}

To summarize, the reported distances to SNRs still have large spreads or uncertainties in many cases. Considering the ambiguity in the kinematic distance, and the scattering of the empirical relation in determining radio-surface-brightness distances, more reliable and accurate distances are needed.
Our extinction distances are derived independently with high accuracy and do not suffer from the near/far distance ambiguity.
We compare our derived extinction distances with the radio-surface-brightness distances from \citet{Pavlovic2013ApJS..204....4P}, and the distances determined by other methods collected from \citet{2020arXiv200307576S}, \citet{Green2019JApA...40...36G}, \citet{2019AJ....157..123L}, \citet{2019ApJ...884..113S}, \citet{2019SerAJ.199...23S}, \citet{Ranasinghe2018MNRAS.477.2243R}, \citet{Shan_2018}, \citet{Kilpatrick2016ApJ...816....1K}, \citet{2010ApJ...725..931M}, \citet{2009ApJ...694L..16H},  \citet{Jackson2008ApJ...674..936J}, and \citet{1999ApJ...525..357S}.
These distances are mainly kinematic distances based on HI/CO observations, as well as distances implied by molecular clouds association, X-ray observations, and optical absorption.
We list the collected distances from literature in Table \ref{UKIDSStable} and Table~\ref{VVVtable} as $d_{\rm rad}$, and $d_{\rm other}$, respectively.
Figure~\ref{comp_d} compares the extinction distances $d_{\rm ext}$ (this work) with (a) the radio-surface-brightness distances $d_{\rm rad}$, and (b) the other known distances $d_{\rm other}$, where we assume these distances have 20\% uncertainties.

\subsubsection{Comparison with the Radio-surface-brightness Distances}

In our sample, there are 36 SNRs with radio-surface-brightness distances from \citet{Pavlovic2013ApJS..204....4P}, including 23 in Group A, 4 in Group B, and 9 in Group C.
Figure~\ref{comp_d} (a) compares the derived extinction distance $d_{\rm ext}$ with the radio-surface-brightness distance $d_{\rm rad}$ for 27 SNRs in Group A (red dots) and B (blue asterisks).  The objects can be divided into three classes:
\begin{itemize}
  \item thirteen of them are consistent with each other, consisting of eleven in group A and two in group B;
  \item six of them have a difference around 1.5 kpc;
  \item eight of them have $d_{\rm ext}$ inconsistent with $d_{\rm rad}$, which are six SNRs in Group A  (G54.4-0.3, G3.8+0.3, G6.1+1.2, G301.4-1.0, G315.9+0.0, and G351.7+0.8) and two SNRs in Group B (G49.2-0.7 and G8.7-0.1). The differences are around 2\,kpc for G3.8+0.3, G301.4-1.0, and G351.7+0.8, which may be considered to be consistent with $d_{\rm ext}$ if the large uncertainties in $d_{\rm rad}$ is taken into account.  The radio-surface-brightness distances of SNR G6.1+1.2 with 6.7 kpc and SNR G315.9+0.0 with 8.2 kpc are beyond the measurable range of our method. G54.4-0.3, G49.2-0.7, and G8.7-0.1 are interacting with adjacent molecular clouds \citep{2010ApJ...712.1147J}, for which the extinction distances agree with their reported kinematic distances \citep{2019AJ....157..123L, Ranasinghe2018MNRAS.477.2243R, 2009ApJ...694L..16H}, which implies that $d_{\rm rad}$ is wrong.
\end{itemize}

\subsubsection{Comparison with the other Distances}

Among our 63 SNRs, 5 SNRs are assigned the range of distance and 18 SNRs have specified distance in \citet{Green2019JApA...40...36G}.
Adding the distances from other literature, a total of 29 SNRs (12, 8 and 9 in Group A, B, C respectively) have specified $d_{\rm other}$ distances which are listed in the column of $d_{\rm other}$ of Table 1 and 2 and marked by different symbols.
Figure~\ref{comp_d} (b) compares the derived extinction distance $d_{\rm ext}$ with the known distance $d_{\rm other}$ for 20 SNRs in Group A (red dots) and B (blue asterisks). It can be seen that the dispersion is apparently smaller than in Figure~\ref{comp_d} (a) for $d_{\rm rad}$. They are divided into three classes as following:
\begin{itemize}
  \item Fifteen of them are consistent with each other, consisting of seven in Group A and eight in Group B including two special cases G359.1-0.5 and G347.3-0.5. As mentioned above, the reported distances to G359.1-0.5 and G347.3-0.5 are controversial in the literature. For G359.1-0.5, our derived distance $3.29\pm0.47$ kpc confirmed that this SNR is in the foreground of the Galactic center at a distance of $\sim$4 kpc by \citet{2020arXiv200307576S}. For G347.3-0.5, our extinction distance of $4.56\pm0.58$ kpc is consistent with the suggested distance of $6\pm1$ kpc by \citet{1999ApJ...525..357S};
  \item Two SNRs have a difference around 1.5 kpc between the two distances;
  \item Three SNRs in Group A (G65.1+0.6, G73.9+0.9, and G351.7+0.8) have the extinction distance inconsistent with the kinematic one. The disagreement comes from both sides. For G65.1+0.6 and G351.7+0.8, the suggested kinematic distances are 9 kpc and 13 kpc, respectively, which are beyond currently measurable ranges of the extinction distances. On the other side, the kinematic distance suffers the usual ambiguity. Besides this, non-circular motion might cause errors. However, the extinction distances of these three SNRs agrees with the radio-surface-brightness distances. Considering that these objects are classified into Group A, the extinction distance should be more reliable than the kinematic distance.
\end{itemize}

\section{Discussion}\label{Discussion}
\subsection{SNRs in the Inner Disk}\label{SNRs_distribution}

\begin{figure}
\centering
\includegraphics[width=\hsize]{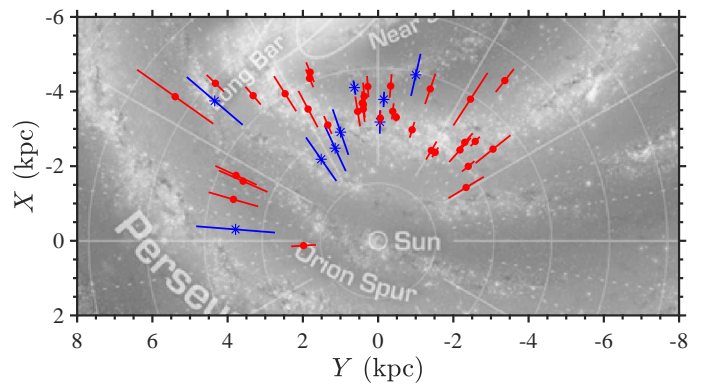}
\caption{The distribution of the 43 SNRs consisted of 34 in Group A (red dots) and 9 in Group B (blue asterisks), respectively in the inner disk superimposed on Robert Hurt's sketch of our Galaxy.}
\label{FigSNRmap}
\end{figure}

The spatial distribution of our 43 SNRs (except 20 SNRs in Group C) in the inner disk is shown in Figure~\ref{FigSNRmap}.
Most of them are located from 2 kpc to 5 kpc away.
It can be seen that some SNRs are associated with the spiral arms, which is expected because massive stars are born in spiral arms.
However, for SNRs in the direction to spiral arms, more than one peak may be found in the extinction gradient.
In our sample, 12 SNRs have two peaks in the extinction gradients in the sightline. For each of them, we carefully scrutinize the distance to the SNR.
Additional information, such as the known distances of $d_{\rm kin}$ and $d_{\rm rad}$, radio observations of SNRs, is used to help identify the distance to the SNR.
For example, there are two jumps in the extinction gradient for G309.8+0.0 (Figure~\ref{FigVVVdAk-1}). We tend to think that the primary sharp one is caused by the SNR, while the broad one may associate with the molecular clouds in the Scutum-Centaurus arm, considering that a SNR should be smaller than a giant molecular cloud in the spiral arm.
The distance of 3.6 kpc determined from the associated molecular cloud \citep{Case1998ApJ...504..761C} agrees with our primary extinction distance.
Moreover, the primary distance is consistent with $d_{\rm rad}$, which further supports our judgment. There are a few objects in-between the spiral arms, which may be caused by Type Ia SNe. Although Type Ia SNs are believed to produce little dust themselves, the energetic explosion should be able to compress the ambient ISM to cause an extinction jump relative to the surrounding area.

\subsection{The Relation of Diameter with Extinction}\label{A-Drelation}

The diameter of a SNR is an indicator of its age. The SNR expands almost homogeneously after  the explosion, and the diameter is then positively correlated to the age. The expansion leads to a more and more tenuous cloud with lower and lower density. In general, the dust density of the SNRs should decrease with age.
Therefore, an inverse relation is expected between the extinction $A_\Ks$ and the diameter $D$ for SNRs.

\begin{figure}
\centering
\includegraphics[width=\hsize]{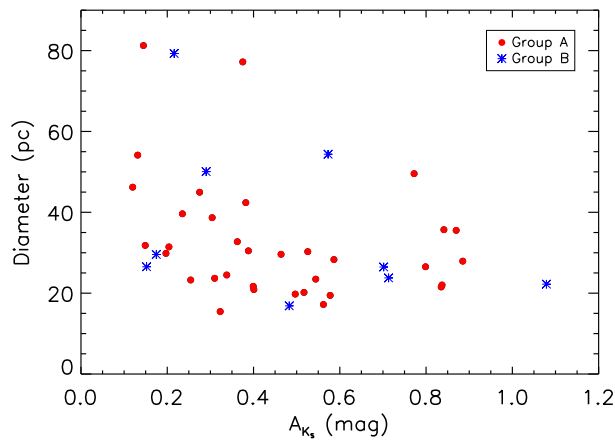}
\caption{The diameter $D$ vs. extinction $A_\Ks$ for the 43 SNRs consisted of 34 in Group A (red dots) and 9 in Group B (blue asterisks), respectively.}
\label{FigSNRAk-D}
\end{figure}

The linear size of a SNR can be calculated from its angular diameter once the distance is known.
The following discussions are based on the 43 SNRs with reliable distances in Group A and B.
With the distance derived above, the linear diameter $D$ is calculated from the angular radius of spherical SNR or the major axis of elliptical SNR.
The calculated diameter of the SNRs ranges from about 15 pc to 80 pc, as shown in Figure \ref{FigSNRAk-D}.
\citet{Draine2011book} estimated the radius of a SNR under typical conditions to be from $\sim$ 5 pc to 24 pc during the Sedov-Taylor phase, and $\sim$ 70 pc when the SNR fade away. According to this model calculation, 36 SNRs are in the Sedov-Taylor phase with a radius smaller than 24 pc, and only seven SNRs should be in the snowplow phase with a radius larger than 24 pc. No one is in the fade-away phase as expected because of the very tenuous structure and thus little additional extinction in this stage.
The SNRs in the free-expansion phase are not included because we chose only big SNRs (larger than $20\arcmin$) that have numerically enough stars to trace the extinction variation. The sizes of the SNRs confirm the distances are reasonable.

The extinction of a SNR can be measured by the jump at the position of the SNR in the distance--extinction diagram, e.g., Figure~\ref{FigD-A}. As the sharp extinction jump is caused by the SNR, the increment of extinction is the extinction of the SNR.
The extinction values of SNRs, ($A_\Ks$)$_{\rm SNR}$, are listed in Table 1 and Table 2.
For example, the increment of extinction $A_\Ks$ for G22.7-0.2 is 0.84 mag at the SNR position (i.e., 4.72 kpc away).
Figure \ref{FigSNRAk-D} displays the distribution of diameter $D$ with extinction $A_\Ks$.
It can be seen that the diameter of the SNR decreases with extinction, with a moderate Pearson correlation coefficient of -0.34.
This consistency with expectation indirectly proves the correctness of our distances.

\subsection{Dust Mass of SNRs}

As the extinction is proportional to the dust column density,
the dust mass of a SNR can be estimated from its extinction,
with the geometry information of a SNR and the optical property of the dust.
\citet{Zhao2018ApJ...855...12Z} estimated the dust mass of the Monoceros SNR using this idea.
In this work, we follow their way to estimate the dust mass of SNRs.
First, we assume the extinction of SNRs complies with the dust model for the Galactic average extinction law at $R_{\rm V}=3.1$ from \citet{WD01}, although this may not be true due to the influence of the SN explosion on the dust. The mass extinction coefficient in the $V$ band is then
\begin{equation}\label{equ2}
K_{{\rm ext}, V} = A_V/\sum\nolimits_{\rm dust} = 2.8\times10^4\ \rm {mag\ cm^2\ g^{-1}}  ~~~,
\end{equation}
where $\sum_{\rm dust}$ is the surface mass density.
With the effective surface area $A_{\rm eff}$, the dust mass is then
$M_{\rm dust}=\sum_{\rm dust} \times A_{\rm eff}$.
\citet{Owen_2015} used a clumped-shell geometry to describe the morphology of SNRs.
For simplicity, we assume a dusty circular shell with $R_{\rm in}$ (the inner radius of the dust shell) and $R_{\rm out}$ (the outer radius of the dust shell).
Therefore, the effective surface area is $A_{\rm eff}={\rm \pi}\ (R_{\rm out}^2 - R_{\rm in}^2) \times F_{\rm fil}$, where $F_{\rm fil}$ is the filling factor.
Consequently, the mass of the dust $M_{\rm dust}$ in the shell can be calculated by
\begin{equation}\label{equ3}
M_{\rm dust}=\frac{A_V \times {\rm \pi}\ (R_{\rm out}^2 - R_{\rm in}^2) \times F_{\rm fil}} {K_{{\rm ext}, V}}  ~~~.
\end{equation}
The average extinction of SNRs in $\Ks$-band has been derived in Section \ref{A-Drelation}, which can be converted to $A_V=A_\Ks/0.11$ by adopting the extinction coefficient for $R_{\rm V}=3.1$ from \citet{WD01}.
Assuming the filling factor $F_{\rm fil}$ to be 0.1 as \citet{Owen_2015} suggested and substituting the coefficients with correct units,
the mass of dust is
\begin{equation}\label{equ4}
M_{\rm dust}=0.488 \frac{A_\Ks}{\rm mag}[(\frac{R_{\rm out}}{\rm pc})^2 - (\frac{R_{\rm in}}{\rm pc})^2] M_\odot ~~~.
\end{equation}

\begin{figure}
\centering
\includegraphics[width=\hsize]{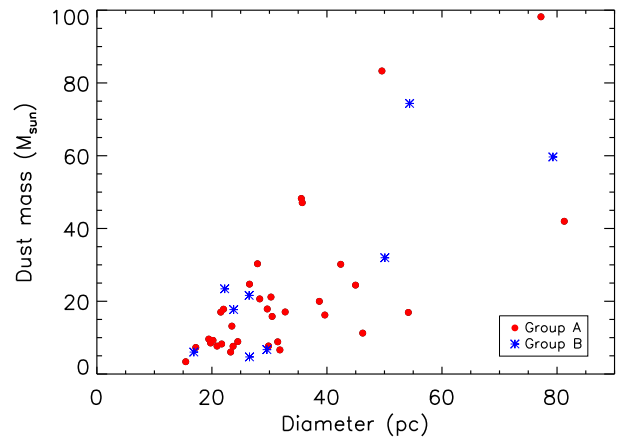}
\caption{The distribution of dust mass vs. diameter for 43 SNRs consisted of 34 in Group A (red dots) and 9 in Group B (blue asterisks), respectively.
}
\label{FigSNRdust}
\end{figure}

Assuming $R_{\rm in}=0.8R_{\rm out}$ that is the same as \citet{Zhao2018ApJ...855...12Z}, the dust mass of each SNR can be deduced by equation~(\ref{equ4}).
Figure~\ref{FigSNRdust} shows the distribution of dust mass versus diameter for 43 SNRs in Group A and B.
Due to the limit of the method, the diameter of sample SNRs starts from about 15 pc and reaches up to 80 pc, the corresponding dust mass ranges from about 2 $M_\odot$ to 100 $M_\odot$.
Generally, SNRs with larger diameter have swept a larger interstellar area and led to more interstellar dust clumped in the shell of SNRs as seen in Figure~\ref{FigSNRdust}.
Among these SNRs, most of them have dust mass less than 30 $M_\odot$.
This result is consistent with previous works.
For example, \citet{Draine2009ASPC..414..453D} estimated the dust mass shocked at the end of the Sedov-Taylor phase to be $\sim$ 10 $M_\odot$ with the corresponding radius of $\sim$ 20 pc. Our estimation of dust mass for the SNRs with a diameter of $\sim$ 20 pc is approximately 10 $M_\odot$. \citet{2015ApJ...799...50L} estimated the dust mass removed by a SNR is about 4 $M_\odot$ in the LMC, which corresponds to about 15 $M_\odot$ per SNR in the Galaxy if the difference of gas-to-dust ratio is taken into account.
There are seven SNRs whose dust mass is over 40 $M_\odot$.
On the one hand, this large mass may be caused by the large size of the SNR. On the other hand, except G318.2+0.1, the remaining six SNRs are associated with molecular clouds where more dust is accumulated due to a dense environment. 

%

Several factors affect the estimation of dust mass from Equation~(\ref{equ3}), such as the extinction coefficient, and the mass extinction coefficient of the dust grain model.
In the previous calculation, we take these parameters from \citet{WD01}.
The dust mass would increase when adopting the extinction coefficient $A_\Ks/A_V=0.078$ from \citet{Wang_Chen2019}, while the dust mass will decrease by taking the mass extinction coefficient $K_{{\rm ext}, V}=3.7\times10^4\ \rm {mag\ cm^2\ g^{-1}}$ from \citet{2013ApJ...770...27N}.
These changes in dust mass are taken as errors and listed in the last columns of Table~\ref{UKIDSStable} and~\ref{VVVtable}.
Note that these dust masses are very crude estimations.
The uncertainty of the inner radius of the dust shell $R_{\rm in}$ and the filling factor $F_{\rm fil}$ can also lead to errors in the dust mass.  Nevertheless, the calculated dust mass reflects the approximate dust mass of SNRs and more accurate calculation is planned.

\subsection{The Extinction Distance}

The extinction distance of a target measures the position of the jump in extinction towards the sightline of the target.
This method was used to independently measure the distance to molecular clouds or SNRs in recent years.
Stead \& Hoare (2010) determined the extinction distances to a group of molecular clouds in the Galactic plane using the near-IR UKIDSS survey.
They stated that the independent measurement of the extinction distances is useful to resolve the kinematic distance ambiguity.
\citet{Foster2012ApJ...751..157F} used near-IR extinction to estimate the distances to dark clouds in the first Galactic quadrant and found that the extinction distances reproduce the maser parallax distances better than the kinematic distances.
\citet{Schlafly_2014} estimated the extinction distance  to a large group of high-latitude molecular clouds, as well as many other well-studied clouds, including Orion, Perseus, Taurus, Cepheus, Polaris, California, and Monoceros R2.
These distances span from 0.1 kpc to 2.4 kpc.
Their technique is limited by the distance up to 5 kpc due to the high interstellar extinction in the optical photometry they used from PanSTARRS-1.
With {\it Gaia} parallaxes, \citet{Zucker_2019} updated the distances to molecular clouds in \citet{Schlafly_2014}.
\citet{Chenbq2017} used three-dimensional (3D) dust extinction map to estimate the distance of SNR S147, and analyzed the SNR interacting with the molecular cloud.
Later, \citet{2019MNRAS.488.3129Y} applied this technique to measure distances of 12 SNRs in the direction of the Galactic anticentre.
They derived accurate distances to four SNRs, and rough estimations of distances to three SNRs.
By measuring the position of the sharp increase of the interstellar extinction in the target sightline, \citet{Zhao2018ApJ...855...12Z} derived the distance to the Monoceros SNR and its two nearby nebulae.
Further, they used this technique to investigate the distances and extinction curves of 32 SNRs, and determined the distances to 15 of them from the extinction they produced and their association with molecular clouds \citep{Zhao2020}.
\citet{Shan_2018, Shan_2019} constructed the optical extinction--distance relations along the directions of SNRs in the first and fourth Galactic quadrant.
With the known extinction of SNRs, the distances or the limits of distances are obtained.

Although the data and the specific procedures used in these works are different, the determined distances are generally consistent.
For example, the distances obtained for 7 SNRs in \citet{2019MNRAS.488.3129Y} are all highly consistent with the results determined in \citet{Zhao2020}.
Two SNRs in our sample are also studied by \citet{Zhao2020}.
One is G93.7-0.2, whose distance (1.99 kpc) in this work is consistent with that of 2.16 kpc of \citet{Zhao2020}.
The other is G65.1+0.6 with a distance of 4.16 kpc in this work, discrepant with 1.33 kpc by \citet{Zhao2020}.
As mentioned in \citet{Zhao2020}, the extinction jump at 1.3 kpc might be caused by a local foreground molecular cloud, which made them classify this result into Level C.
Meanwhile, this object is classified into Level A in this work due to its prominent extinction and gradient in comparison with the surrounding area so that the result is highly reliable.
Although our derived extinction distance deviates from the suggested kinematic distance of 9 kpc by \citet{2006A&A...455.1053T}, it is consistent with the radio-surface-brightness distance of 2.6 kpc by \citet{Pavlovic2013ApJS..204....4P}.
It is worth noting that  \citet{2019MNRAS.488.3129Y} and \citet{Zhao2020} can only investigate SNRs within 4 kpc limited by the accuracy of parallaxes from {\it Gaia}, while our technique can measure much farther SNRs, with some cases between 4 and 7 kpc and a couple of extreme cases such as G55.0+0.3 at 10.18 kpc and G36.6-0.7 at 8.66 kpc.

In our sample, five SNRs are reported in \citet{Shan_2018}.
The agreement is found in three SNRs, G34.7-0.4, G49.2-0.7, G85.4+0.7, all in Group B.
For the other two SNRs (G18.9-1.1 and G66.0+0.0), the difference is apparent, specifically, the distances from \citet{Shan_2018} are about 2 kpc, which are smaller than ours.
In order to find the reason, we re-calculated the distance of all the five SNRs with our distance--extinction diagrams, but at the position of the extinction adopted by \citet{Shan_2018}, i.e. the same way as \citet{Shan_2018} did. It is found that the distance to G18.9-1.1 (Group B) becomes 3.7 kpc, which is apparently larger than 1.8 kpc by \citet{Shan_2018} and consistent with our result in Table 1 (3.08 kpc). While for G66.0+0.0 (Group A), the result (2.0 kpc) is smaller than our result in Table 1 (3.93 kpc) and consistent with \citet{Shan_2018} (2.3 kpc). It seems the discrepancy can be caused by the method both to determine the extinction and to determine the distance. 
More optical data are needed to clarify this discrepancy.
Although both our work and Shan et al.'s work took RCs as tracers to measure distances of SNRs by using near-IR photometric data, our method has the advantages of deeper photometry by adding the VVV and UKIDSS data, cleaner sample by excluding some dwarfs and giants with the {\it Gaia} data, and more accurate method to derive the distance.
Nevertheless, as mentioned in Section~\ref{Result}, the lower limit of the measurable distance is about 2 kpc if we use RCs as the tracers and work in the near-infrared bands.
Instead, optical data for RCs or taking dwarfs as tracers can extend the lower limit of extinction distance.

\section{Conclusion}\label{Conclusion}

The distance of a SNR is an essential and important parameter.
However, the determination is hard and tough.
In this work, we determine the distances to a large sample of SNRs in the inner disk by taking RC stars as the tracers of the distance and extinction.
The main results of this work are as follows:

\begin{enumerate}
\item Based on near-IR photometric data from the 2MASS, and UKIDSS or VVV surveys, we determined the distances to 63 SNRs in the Galactic inner disk according to the additional extinction and extinction gradient relative to the surrounding area. These SNRs are divided into three levels, namely 34, 9, 20 ones in Level A, B, and C, respectively, with decreasing reliability of distances.
The distances in Level A are highly reliable, while SNRs in Level B are associated with molecular clouds and their distances are reliable as well.
The distances in Level C  needs further investigations, but they can be safely taken as reference. 
The distance is compared with those derived from the radio brightness or the associated molecular cloud. 
Our results are consistent in many cases, which helps to clarify some discrepant results with higher confidence.
Accurate distances to seven SNRs, G5.4-1.2, G308.8-0.1, G318.2+0.1, G318.9+0.4, G327.1-1.1, G329.7+0.4, and G341.2+0.9, are determined for the first time.
Through elaborate comparison, we conclude that the measurement of extinction distances is a convenient and reliable method.
In the future, with deeper IR photometric data, we can extend the maximum measurable distance of SNRs.
\item With the derived distances and extinction of SNRs, we calculated the diameters and the dust mass of SNRs.
We find an inverse relation between the extinction and the diameter as expected. The calculated dust mass conforms to the evolution model of an SNR. These results indirectly prove the correctness of our determined distance and extinction.
\item It is worth noting that if there are other obvious dusty sources in the SNR sightline, more than one jump in extinction will be found. In such a case, we combine with other methods to find which extinction jump is caused by the SNR and the corresponding distance to the SNR. In other words, this method can be applied to measure the distances to any extended dusty sources, such as molecular clouds. In the future, we will use it to determine distances of molecular clouds in the spiral arms and study of the Galactic structure.
\end{enumerate}

\begin{acknowledgements}
We thank the anonymous referee for very useful suggestions to improve the quality and readability of the paper.
We thank Dr. Jiaming Liu for very helpful discussion.
This work is supported by the National Natural Science Foundation of China through the projects NSFC 11533002, U1631104, and 11933004. 
X.C. acknowledges support by the NSFC project 11903045.
This work has made use of data from the surveys by {\it Gaia}, 2MASS, UKIDSS, and VVV.
This work has made use of data from the European Space Agency (ESA) mission {\it Gaia} (https://www.cosmos.esa.int/gaia), processed by the {\it Gaia} Data Processing and Analysis Consortium (DPAC, https://www.cosmos.esa.int/web/gaia/dpac/consortium). Funding for the DPAC has been provided by national institutions, in particular the institutions participating in the {\it Gaia} Multilateral Agreement.
The Two Micron All Sky Survey is a joint project of the University of Massachusetts and the Infrared Processing and Analysis Center/California Institute of Technology, funded by the NASA and the NSF.
We gratefully acknowledge use of data from the ESO Public Survey program ID 179.B-2002 taken with the VISTA telescope, data products from the Cambridge Astronomical Survey Unit.
\end{acknowledgements}

\bibliographystyle{aa} 
\bibliography{reference} 

\clearpage
\begin{sidewaystable*}
\caption{The distance and extinction of 35 SNRs measured by UKIDSS data.}
\label{UKIDSStable}
\scriptsize
\centering
\begin{tabular}{llclcccccccclcl}
\hline\hline
Name\tablefootmark{a} & Other Name & R.A. & Dec. & Radius & RC Ridge\tablefootmark{b}
& $(d_{\rm ext})_{\rm P}$\tablefootmark{c} & $\rm (\Delta A_\Ks)_P$
& $(d_{\rm ext})_{\rm S}$\tablefootmark{d}  & $\rm (\Delta A_\Ks)_S$
& Reliability & $d_{\rm rad}$\tablefootmark{e}  & $d_{\rm other}$\tablefootmark{f}
& ($A_\Ks$)$_{\rm SNR}$ & Dust Mass \\
  & & (deg) & (deg) & (deg) &
& (kpc) & (mag/kpc) & (kpc) & (mag/kpc) &  & (kpc) & (kpc)
& (mag) & (M$_\odot$) \\
\hline
G5.4-1.2$^a$	&	Milne 56	&	270.54	&	-24.90	&	0.29	&	Normal	&	$\bm{3.89 \pm0.91}$	&	0.15	&	0	&	0	&	A	&		&	>4.3	&	0.275 	&	$18.94^{+7.10 }_{-3.34 }$	\\
G6.1+1.2	&		&	268.73	&	-23.08	&	0.25	&	Normal	&	$\bm{3.27 \pm0.73}$	&	0.10	&	0	&	0	&	A	&	6.7 	&		&	0.138 	&	$4.27 ^{+1.60 }_{-0.75 }$	\\
G6.4-0.1$^a$	&	W28	&	270.13	&	-23.43	&	0.40	&	KDE	&	$\bm{3.55 \pm0.90}$	&	0.40	&	0	&	0	&	A	&		&	1.9	&	0.742 	&	$79.96^{+29.98}_{-14.11}$	\\
G8.9+0.4	&		&	270.99	&	-21.05	&	0.20	&	KDE	&	$\bm{3.54 \pm0.62}$	&	0.27	&	0	&	0	&	A	&	4.3 	&		&	0.376 	&	$10.06^{+3.77 }_{-1.78 }$	\\
G13.3-1.3	&		&	274.83	&	-18.00	&	0.58	&	Normal	&	$\bm{4.76 \pm0.93}$	&	0.10	&	0	&	0	&	C	&		&	2.0--4.0	&	0.214 	&	$50.65^{+18.99}_{-8.94 }$	\\
G15.1-1.6	&		&	276.00	&	-16.57	&	0.25	&	Normal	&	$\bm{2.91 \pm0.68}$	&	0.16	&	0	&	0	&	C	&	4.5 	&	2.2\tablefootmark{$\lhd$}	&	0.186 	&	$4.21 ^{+1.58 }_{-0.74 }$	\\
G18.9-1.1$^a$	&		&	277.46	&	-12.97	&	0.28	&	Normal	&	$5.47 \pm0.79$	&	0.25	&	$\bm{3.08\pm0.65}$	&	0.13	&	B	&		&	1.8\tablefootmark{$\star$}/2.0\tablefootmark{$\diamond$}	&	0.175 	&	$6.70 ^{+2.51 }_{-1.18 }$	\\
G19.1+0.2	&		&	276.23	&	-12.12	&	0.23	&	Normal	&	$\bm{3.57 \pm0.67}$	&	0.15	&	0	&	0	&	C	&	4.0 	&		&	0.269 	&	$9.29 ^{+3.48 }_{-1.64 }$	\\
G21.8-0.6$^a$	&	Kes 69	&	278.19	&	-10.13	&	0.17	&	Normal	&	$\bm{4.87 \pm0.29}$	&	0.81	&	$3.56\pm0.24$	&	0.4	&	A	&	3.2 	&	5.2\tablefootmark{$\diamond$}/5.6\tablefootmark{$\bullet$}	&	0.586 	&	$20.64^{+7.74 }_{-3.64 }$	\\
G22.7-0.2$^a$	&		&	278.31	&	-9.22	&	0.22	&	Normal	&	$\bm{4.72 \pm0.26}$	&	1.33	&	$3.11\pm0.29$	&	0.49	&	A	&	3.2 	&	4.4/4.7\tablefootmark{$\bullet$}	&	0.841 	&	$47.09^{+17.66}_{-8.31 }$	\\
G23.3-0.3$^a$	&	W41	&	278.69	&	-8.80	&	0.23	&	Normal	&	$\bm{3.38 \pm0.26}$	&	1.34	&	$4.14\pm0.27$	&	0.93	&	A	&	2.7 	&	4.2\tablefootmark{$\diamond$}/4.8\tablefootmark{$\bullet$}	&	0.799 	&	$24.70^{+9.26 }_{-4.36 }$	\\
G24.7+0.6$^a$	&		&	278.54	&	-7.08	&	0.25	&	Normal	&	$\bm{2.73 \pm0.68}$	&	0.31	&	$5.87\pm0.71$	&	0.26	&	B	&		&	3.5\tablefootmark{$\diamond$}	&	0.483 	&	$6.03 ^{+2.26 }_{-1.06 }$	\\
G25.1-2.3	&		&	281.29	&	-8.00	&	0.67	&	Normal	&	$\bm{3.45 \pm0.83}$	&	0.05	&	0	&	0	&	C	&		&	2.9	&	0.100 	&	$10.66^{+4.00 }_{-1.88 }$	\\
G27.8+0.6$^a$	&		&	279.96	&	-4.40	&	0.42	&	KDE	&	$\bm{3.99 \pm0.55}$	&	0.21	&	0	&	0	&	A	&		&	2--3\tablefootmark{$\oslash$}	&	0.275 	&	$24.43^{+9.16 }_{-4.31 }$	\\
G30.7+1.0	&		&	281.00	&	-1.53	&	0.20	&	Normal	&	$\bm{3.64 \pm0.93}$	&	0.13	&	0	&	0	&	C	&	5.1 	&		&	0.242 	&	$5.16 ^{+1.93 }_{-0.91 }$	\\
G32.1-0.9$^a$	&		&	283.29	&	-1.13	&	0.33	&	KDE	&	$\bm{4.65 \pm0.56}$	&	0.11	&	0	&	0	&	A	&		&	4.6\tablefootmark{$\diamond$}	&	0.131 	&	$16.93^{+6.35 }_{-2.99 }$	\\
G34.7-0.4$^a$	&	W44, 3C392	&	284.00	&	1.37	&	0.29	&	Normal	&	$\bm{2.66 \pm0.71}$	&	0.49	&	0	&	0	&	B	&		&	2.1\tablefootmark{$\star$}/2.8\tablefootmark{$\diamond$}/3.0\tablefootmark{$\bullet$}	&	0.713 	&	$17.71^{+6.64 }_{-3.12 }$	\\
G36.6-0.7	&		&	285.15	&	2.93	&	0.21	&	Normal	&	$\bm{8.66 \pm1.17}$	&	0.10	&	0	&	0	&	C	&		&		&	0.291 	&	$50.75^{+19.03}_{-8.96 }$	\\
G38.7-1.3$^a$	&		&	286.67	&	4.47	&	0.27	&	KDE	&	$\bm{4.11 \pm0.88}$	&	0.08	&	0	&	0	&	C	&		&		&	0.142 	&	$5.44 ^{+2.04 }_{-0.96 }$	\\
G40.5-0.5$^a$	&		&	286.79	&	6.52	&	0.18	&	KDE	&	$\bm{5.12 \pm0.32}$	&	0.44	&	0	&	0	&	A	&	4.4 	&		&	0.362 	&	$17.06^{+6.40 }_{-3.01 }$	\\
G42.8+0.6	&		&	286.83	&	9.08	&	0.20	&	KDE	&	$\bm{4.24 \pm0.93}$	&	0.13	&	0	&	0	&	C	&	5.4 	&		&	0.251 	&	$9.67 ^{+3.63 }_{-1.71 }$	\\
G43.9+1.6	&		&	286.46	&	10.50	&	0.50	&	KDE	&	$\bm{1.52 \pm0.60}$	&	0.07	&	$5.56\pm0.53$	&	0.05	&	C	&	2.5 	&		&	0.093 	&	$2.88 ^{+1.08 }_{-0.51 }$	\\
G45.7-0.4	&		&	289.10	&	11.15	&	0.18	&	Normal	&	$\bm{6.04 \pm0.33}$	&	0.31	&	0	&	0	&	A	&	5.3 	&		&	0.304 	&	$19.98^{+7.49 }_{-3.53 }$	\\
G49.2-0.7$^a$	&	W51	&	290.96	&	14.10	&	0.25	&	Normal	&	$\bm{5.74 \pm0.98}$	&	0.14	&	0	&	0	&	B	&	2.1 	&	5.4\tablefootmark{$\bullet$}/5.7\tablefootmark{$\star$}/6\tablefootmark{$\diamond$}	&	0.290 	&	$31.96^{+11.98}_{-5.64 }$	\\
G54.4-0.3$^a$	&	HC40	&	293.33	&	18.93	&	0.33	&	Normal	&	$\bm{6.64 \pm1.25}$	&	0.13	&	$2.4 \pm0.63$	&	0.1	&	A	&	2.5 	&	6.6\tablefootmark{$\diamond$}	&	0.375 	&	$98.18^{+36.82}_{-17.33}$	\\
G55.0+0.3	&		&	293.00	&	19.83	&	0.17	&	Normal	&	$\bm{10.18\pm1.28}$	&	0.09	&	$6.7 \pm1.0$	&	0.08	&	C	&	9.4 	&	14	&	0.262 	&	$30.29^{+11.36}_{-5.35 }$	\\
G59.8+1.2	&		&	294.73	&	24.32	&	0.17	&	Normal	&	$\bm{5.43 \pm1.11}$	&	0.06	&	0	&	0	&	C	&		&	7.3\tablefootmark{$\times$}	&	0.139 	&	$4.87 ^{+1.83 }_{-0.86 }$	\\
G65.1+0.6$^a$	&		&	298.67	&	28.58	&	0.75	&	KDE	&	$\bm{4.16 \pm0.61}$	&	0.11	&	0	&	0	&	A	&	2.6 	&	9	&	0.145 	&	$41.98^{+15.74}_{-7.41 }$	\\
G66.0+0.0	&		&	299.46	&	29.05	&	0.26	&	Normal	&	$\bm{3.93 \pm0.71}$	&	0.09	&	0	&	0	&	A	&		&	2.3\tablefootmark{$\star$}	&	0.149 	&	$6.62 ^{+2.48 }_{-1.17 }$	\\
G73.9+0.9$^a$	&		&	303.56	&	36.20	&	0.23	&	Normal	&	$\bm{4.00 \pm0.69}$	&	0.14	&	0	&	0	&	A	&	4.0 	&	1.25\tablefootmark{$\lhd$}	&	0.204 	&	$8.85 ^{+3.32 }_{-1.56 }$	\\
G85.4+0.7$^a$	&		&	312.67	&	45.37	&	0.20	&	Normal	&	$\bm{3.80 \pm1.05}$	&	0.08	&	0	&	0	&	B	&		&	$3.5\pm1.0$\tablefootmark{$\circ$}/4.4\tablefootmark{$\star$}	&	0.152 	&	$4.70 ^{+1.76 }_{-0.83 }$	\\
G85.9-0.6	&		&	314.67	&	44.88	&	0.20	&	Normal	&	$\bm{3.27 \pm0.97}$	&	0.05	&	0	&	0	&	C	&		&	$4.8\pm1.6$\tablefootmark{$\circ$}	&	0.082 	&	$1.88 ^{+0.70 }_{-0.33 }$	\\
G93.7-0.2	&	CTB 104A, DA 551	&	322.33	&	50.83	&	0.67	&	KDE	&	$4.29 \pm0.45$	&	0.17	&	$\bm{1.99\pm0.33}$	&	0.14	&	A	&	1.4 	&	1.5	&	0.120 	&	$11.25^{+4.22 }_{-1.98 }$	\\
G359.0-0.9$^a$	&		&	266.71	&	-30.27	&	0.19	&	Normal	&	$\bm{3.49 \pm0.36}$	&	0.60	&	0	&	0	&	A	&	3.7 	&		&	0.399 	&	$9.55 ^{+3.58 }_{-1.69 }$	\\
G359.1-0.5$^a$	&		&	266.38	&	-29.95	&	0.20	&	Normal	&	$\bm{3.29 \pm0.47}$	&	1.03	&	0	&	0	&	B	&	4.0 	&	8.5\tablefootmark{$\lhd$}/4\tablefootmark{$\bigtriangleup$}	&	1.073 	&	$24.87^{+9.33 }_{-4.39 }$	\\
\hline
\end{tabular}
\tablefoot{\\
\tablefoottext{a}{The known SNR and molecular cloud associations are denoted by "a".}\\
\tablefoottext{b}{The method for obtaining the RC ridge in CMDs: a kernel density estimation (KDE) or a normal parameter estimation (Normal).}\\
\tablefoottext{c}{The primary distance marked as "P".}\\
\tablefoottext{d}{The secondary distance marked as "S".}\\
\tablefoottext{e}{The radio-surface-brightness distances $d_{\rm rad}$ are from \citet{Pavlovic2013ApJS..204....4P}.}\\
\tablefoottext{f}{The distances listed in this column $d_{\rm other}$ are mainly kinematic distances from \citet{Green2019JApA...40...36G}.  
The symbols $\lhd$, $\diamond$, $\bullet$, $\oslash$, $\star$, $\times$, $\circ$, and $\bigtriangleup$ denote distances determined by different methods from \citet{2019SerAJ.199...23S}, \citet{2019AJ....157..123L}, \citet{Ranasinghe2018MNRAS.477.2243R}, \citet{2010ApJ...725..931M}, \citet{Shan_2018}, \citet{Kilpatrick2016ApJ...816....1K}, \citet{Jackson2008ApJ...674..936J}, and \citet{2020arXiv200307576S}, respectively.}\\
}
\end{sidewaystable*}

\clearpage

\begin{sidewaystable*}
\caption{The distance and extinction of 34 SNRs measured by VVV data.}
\label{VVVtable}
\scriptsize
\centering
\begin{tabular}{llclcccccccclcl}
\hline\hline
Name\tablefootmark{a} & Other Name & R.A. & Dec. & Radius & RC Ridge\tablefootmark{b}
& $(d_{\rm ext})_{\rm P}$\tablefootmark{c} & $\rm (\Delta A_\Ks)_P$
& $(d_{\rm ext})_{\rm S}$\tablefootmark{d}  & $\rm (\Delta A_\Ks)_S$
& Reliability & $d_{\rm rad}$\tablefootmark{e}  & $d_{\rm other}$\tablefootmark{f}
& ($A_\Ks$)$_{\rm SNR}$ & Dust Mass \\
  & & (deg) & (deg) & (deg) &
& (kpc) & (mag/kpc) & (kpc) & (mag/kpc) &  & (kpc) & (kpc)
& (mag) & (M$_\odot$) \\
\hline
G3.8+0.3	&		&	268.23	&	-25.47	&	0.15	&	Normal	&	$\bm{4.14 \pm0.29}$	&	0.32	&	0	&	0	&	A	&	6.4	&		&	0.399 	&	$8.25 ^{+3.10 }_{-1.46 }$	\\
G5.4-1.2$^a$	&	Milne 56	&	270.54	&	-24.90	&	0.29	&	Normal	&	$\bm{3.89 \pm0.37}$	&	0.14	&	0	&	0	&	A	&		&	>4.3	&	0.235 	&	$16.22^{+6.08 }_{-2.86 }$	\\
G6.1+1.2	&		&	268.73	&	-23.08	&	0.25	&	Normal	&	$\bm{3.67 \pm0.36}$	&	0.12	&	0	&	0	&	A	&	6.7	&		&	0.197 	&	$7.71 ^{+2.89 }_{-1.36 }$	\\
G6.4-0.1$^a$	&	W28	&	270.13	&	-23.43	&	0.40	&	KDE	&	$\bm{3.55 \pm0.34}$	&	0.56	&	0	&	0	&	A	&		&	1.9	&	0.772 	&	$83.30^{+31.24}_{-14.70}$	\\
G6.5-0.4	&		&	270.55	&	-23.57	&	0.15	&	Normal	&	$\bm{3.72 \pm0.21}$	&	0.70	&	0	&	0	&	A	&	4.1	&		&	0.578 	&	$9.60 ^{+3.60 }_{-1.69 }$	\\
G8.7-0.1$^a$	&	W30	&	271.38	&	-21.43	&	0.38	&	KDE	&	$\bm{4.15 \pm0.19}$	&	0.62	&	0	&	0	&	B	&	1.9	&	4.5\tablefootmark{$\wedge$}	&	0.573 	&	$74.37^{+27.89}_{-13.12}$	\\
G8.9+0.4	&		&	270.99	&	-21.05	&	0.20	&	KDE	&	$\bm{3.51 \pm0.41}$	&	0.21	&	0	&	0	&	A	&	4.3	&		&	0.338 	&	$8.91 ^{+3.34 }_{-1.57 }$	\\
G296.1-0.5	&		&	177.79	&	-62.57	&	0.31	&	KDE	&	$\bm{3.80 \pm0.50}$	&	0.09	&	0	&	0	&	C	&		&	$3.0\pm1.0$\tablefootmark{$\ast$}	&	0.195 	&	$9.68 ^{+3.63 }_{-1.71 }$	\\
G301.4-1.0	&		&	189.48	&	-63.82	&	0.31	&	Normal	&	$\bm{2.74 \pm0.55}$	&	0.12	&	0	&	0	&	A	&	5.2	&		&	0.254 	&	$6.04 ^{+2.27 }_{-1.07 }$	\\
G308.8-0.1	&		&	205.63	&	-62.38	&	0.25	&	Normal	&	$\bm{3.92 \pm0.60}$	&	0.28	&	0	&	0	&	A	&		&	$6.9^{+8.1}_{-2.9}$\tablefootmark{$\ast$}	&	0.885 	&	$30.30^{+11.36}_{-5.35 }$	\\
G309.8+0.0	&		&	207.63	&	-62.08	&	0.21	&	Normal	&	$\bm{3.12 \pm0.22}$	&	0.38	&	$5.61\pm0.42$	&	0.3	&	A	&	4 	&		&	0.497 	&	$8.52 ^{+3.20 }_{-1.50 }$	\\
G312.4-0.4$^a$	&		&	213.25	&	-61.73	&	0.32	&	Normal	&	$\bm{4.41 \pm0.50}$	&	0.25	&	0	&	0	&	C	&	2.4	&	>6/>14/$6.0^{+8.0}_{0.0}$\tablefootmark{$\ast$}	&	0.600 	&	$62.50^{+23.44}_{-11.03}$	\\
G315.4-0.3	&		&	218.98	&	-60.60	&	0.20	&	Normal	&	$\bm{3.31 \pm0.28}$	&	0.26	&	$5.94\pm0.36$	&	0.25	&	C	&		&		&	0.351 	&	$4.48 ^{+1.68 }_{-0.79 }$	\\
G315.9+0.0	&		&	219.60	&	-60.18	&	0.21	&	KDE	&	$\bm{3.71 \pm0.18}$	&	0.61	&	0	&	0	&	A	&	8.2	&		&	0.517 	&	$9.25 ^{+3.47 }_{-1.63 }$	\\
G316.3+0.0	&	MSH 14-57	&	220.38	&	-60.00	&	0.24	&	Normal	&	$\bm{3.84 \pm0.30}$	&	0.55	&	0	&	0	&	C	&	4.1	&	>7.2/$7.2\pm0.6$\tablefootmark{$\ast$}	&	0.810 	&	$18.04^{+6.76 }_{-3.18 }$	\\
G318.2+0.1	&		&	223.71	&	-59.07	&	0.33	&	KDE	&	$\bm{3.27 \pm0.44}$	&	0.45	&	0	&	0	&	A	&		&		&	0.870 	&	$48.25^{+18.10}_{-8.52 }$	\\
G318.9+0.4	&		&	224.63	&	-58.48	&	0.25	&	Normal	&	$\bm{3.50 \pm0.32}$	&	0.28	&	0	&	0	&	A	&		&		&	0.401 	&	$7.68 ^{+2.88 }_{-1.36 }$	\\
G320.4-1.2	&	MSH 15-52, RCW 89	&	228.63	&	-59.13	&	0.29	&	Normal	&	$\bm{3.00 \pm0.45}$	&	0.08	&	$5.85\pm0.22$	&	0.05	&	C	&		&	5.2	&	0.185 	&	$7.58^{+2.84 }_{-1.34 }$	\\
G320.6-1.6	&		&	229.46	&	-59.27	&	0.50	&	KDE	&	$\bm{3.18 \pm0.62}$	&	0.06	&	0	&	0	&	C	&		&		&	0.150 	&	$10.13^{+3.80 }_{-1.79 }$	\\
G321.9-0.3	&		&	230.17	&	-57.57	&	0.26	&	KDE	&	$\bm{5.46 \pm0.39}$	&	0.22	&	0	&	0	&	A	&	3.8	&	$6.5^{+3.5}_{-1.0}$\tablefootmark{$\ast$}	&	0.382 	&	$30.16^{+11.31}_{-5.32 }$	\\
G321.9-1.1	&		&	230.94	&	-58.22	&	0.23	&	Normal	&	$\bm{3.29 \pm0.75}$	&	0.09	&	0	&	0	&	C	&		&		&	0.276 	&	$8.69 ^{+3.26 }_{-1.53 }$	\\
G327.1-1.1	&		&	238.60	&	-55.15	&	0.15	&	Normal	&	$\bm{4.52 \pm0.84}$	&	0.09	&	0	&	0	&	A	&		&		&	0.310 	&	$7.63 ^{+2.86 }_{-1.35 }$	\\
G327.4+0.4	&	Kes 27	&	237.08	&	-53.82	&	0.18	&	Normal	&	$\bm{2.81 \pm0.16}$	&	0.64	&	0	&	0	&	A	&	3.7	&	4.3--5.4	&	0.562 	&	$7.30 ^{+2.74 }_{-1.29 }$	\\
G329.7+0.4	&		&	240.33	&	-52.30	&	0.33	&	Normal	&	$\bm{2.80 \pm0.28}$	&	0.28	&	0	&	0	&	A	&		&		&	0.464 	&	$17.87^{+6.70 }_{-3.15 }$	\\
G335.2+0.1	&		&	246.94	&	-48.78	&	0.18	&	Normal	&	$\bm{3.91 \pm0.49}$	&	0.49	&	0	&	0	&	C	&	4.2	&	1.8\tablefootmark{$\ast$}	&	1.035 	&	$25.95^{+9.73 }_{-4.58 }$	\\
G341.2+0.9	&		&	251.90	&	-43.78	&	0.18	&	KDE	&	$\bm{4.30 \pm0.43}$	&	0.28	&	0	&	0	&	A	&		&		&	0.544 	&	$13.17^{+4.94 }_{-2.32 }$	\\
G343.1-0.7	&		&	255.10	&	-43.23	&	0.23	&	KDE	&	$\bm{3.11 \pm0.22}$	&	0.86	&	0	&	0	&	A	&	4.5	&		&	0.835 	&	$17.02^{+6.38 }_{-3.00 }$	\\
G347.3-0.5$^a$	&	RX J1713.7-3946	&	258.46	&	-39.75	&	0.54	&	KDE	&	$\bm{4.56 \pm0.58}$	&	0.09	&	0	&	0	&	B	&		&	1.3\tablefootmark{$\dagger$}/$6\pm1$\tablefootmark{$\vee$}	&	0.216 	&	$59.66^{+22.37}_{-10.53}$	\\
G351.7+0.8	&		&	260.25	&	-35.45	&	0.15	&	Normal	&	$\bm{3.35 \pm0.11}$	&	0.64	&	0	&	0	&	A	&	5.4	&	$13.2\pm0.5$\tablefootmark{$\ast$}	&	0.323 	&	$3.38 ^{+1.27 }_{-0.60 }$	\\
G353.6-0.7	&		&	263.00	&	-34.73	&	0.25	&	Normal	&	$\bm{3.49 \pm0.22}$	&	0.35	&	0	&	0	&	A	&	4.9	&	3.2\tablefootmark{$+$}	&	0.388 	&	$15.83^{+5.93 }_{-2.79 }$	\\
G355.4+0.7	&		&	262.83	&	-32.43	&	0.21	&	Normal	&	$\bm{4.16 \pm0.32}$	&	0.36	&	0	&	0	&	A	&	4.8	&		&	0.526 	&	$21.16^{+7.93 }_{-3.73 }$	\\
G357.7+0.3$^a$	&		&	264.65	&	-30.73	&	0.20	&	KDE	&	$\bm{3.79 \pm0.21}$	&	0.66	&	0	&	0	&	B	&	4.2	&		&	0.702 	&	$21.61^{+8.10 }_{-3.81 }$	\\
G359.0-0.9$^a$	&		&	266.71	&	-30.27	&	0.19	&	Normal	&	$\bm{3.29 \pm0.20}$	&	0.96	&	0	&	0	&	A	&	3.7	&		&	0.837 	&	$17.82^{+6.68 }_{-3.14 }$	\\
G359.1-0.5$^a$	&		&	266.38	&	-29.95	&	0.20	&	Normal	&	$\bm{3.18 \pm0.32}$	&	0.81	&	0	&	0	&	B	&	4 	&	8.5\tablefootmark{$\lhd$}/4\tablefootmark{$\bigtriangleup$}	&	1.079 	&	$23.42^{+8.78 }_{-4.13 }$	\\
\hline
\end{tabular}
\tablefoot{\\
\tablefoottext{a}{The known SNR and molecular cloud associations are denoted by "a".}\\
\tablefoottext{b}{The method for obtaining the RC ridge in CMDs: a kernel density estimation (KDE) or a normal parameter estimation (Normal).}\\
\tablefoottext{c}{Primary distance marked as "P".}\\
\tablefoottext{d}{Secondary distance marked as "S".}\\
\tablefoottext{e}{The radio-surface-brightness distances $d_{\rm rad}$ are from \citet{Pavlovic2013ApJS..204....4P}.}\\
\tablefoottext{f}{The known distances $d_{\rm other}$ are mainly kinematic distances from \citet{Green2019JApA...40...36G}, where the symbol $\dagger$ describes the distance implied by the associated molecular clouds and X-ray observations, the symbol $+$ describes the distance suggested by various observations.
The symbol $\ast$ denotes distances summarized in \citet{2019ApJ...884..113S}.
The symbols $\wedge$, $\vee$, $\lhd$, $\bigtriangleup$ denote distances from \citet{2009ApJ...694L..16H}, \citet{1999ApJ...525..357S}, \citet{2019SerAJ.199...23S}, \citet{2020arXiv200307576S}, respectively.}\\
}
\end{sidewaystable*}


%
\begin{appendix}
\section{Appendix}

\begin{figure*}
   \centering
   \includegraphics[width=\hsize]{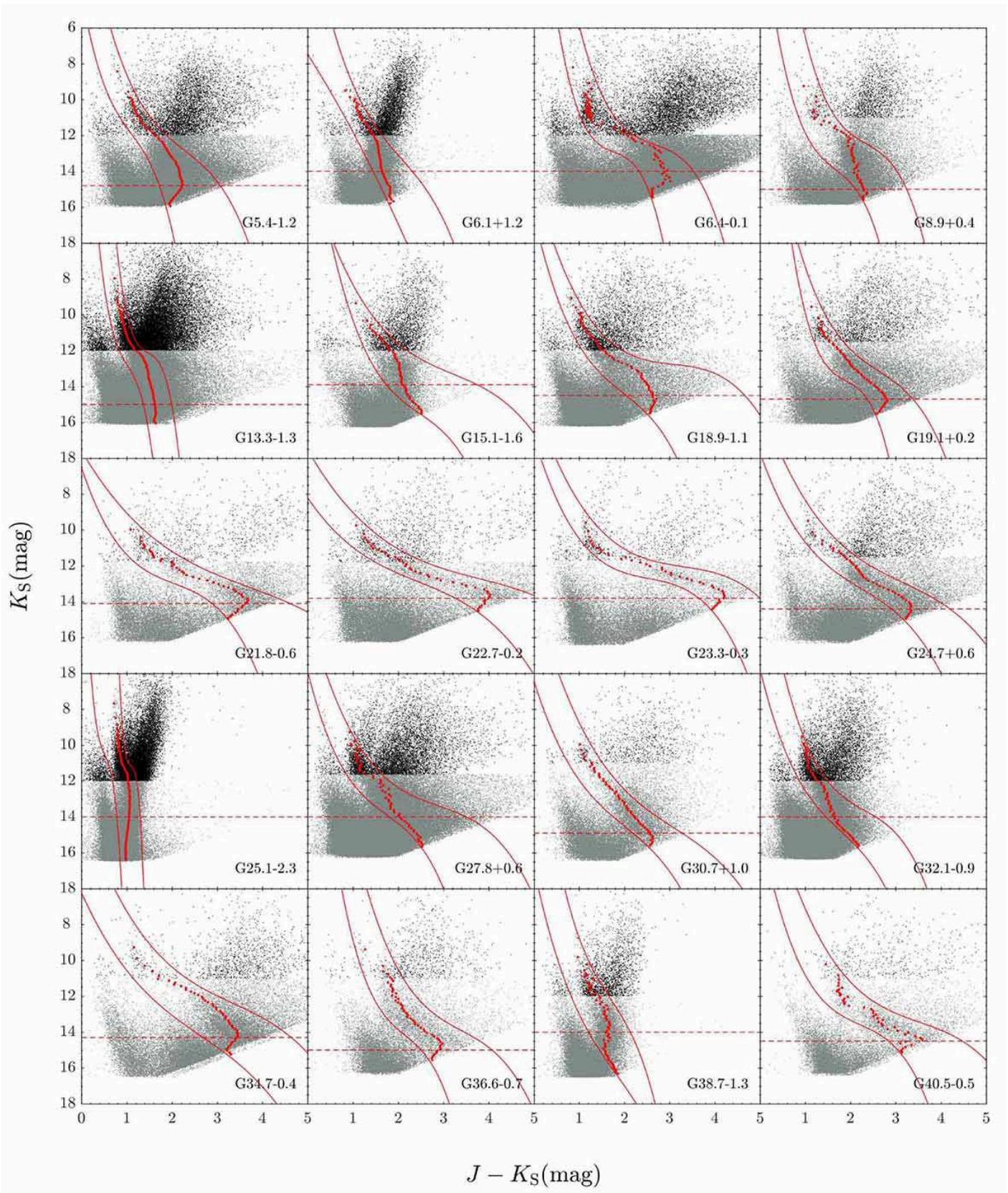}
   \caption{CMDs for SNRs in UKIDSS.
   Black points denote the 2MASS data,
   grey points denote the UKIDSS data,
   red cubic curves roughly outline the range of RCs,
   red points are the RC ridge,
   and red dashed line represents the cutoff magnitude.}
   \label{FigUKICMD-1}
   \end{figure*}

\begin{figure*}[htbp]
   \centering
   \includegraphics[width=\hsize]{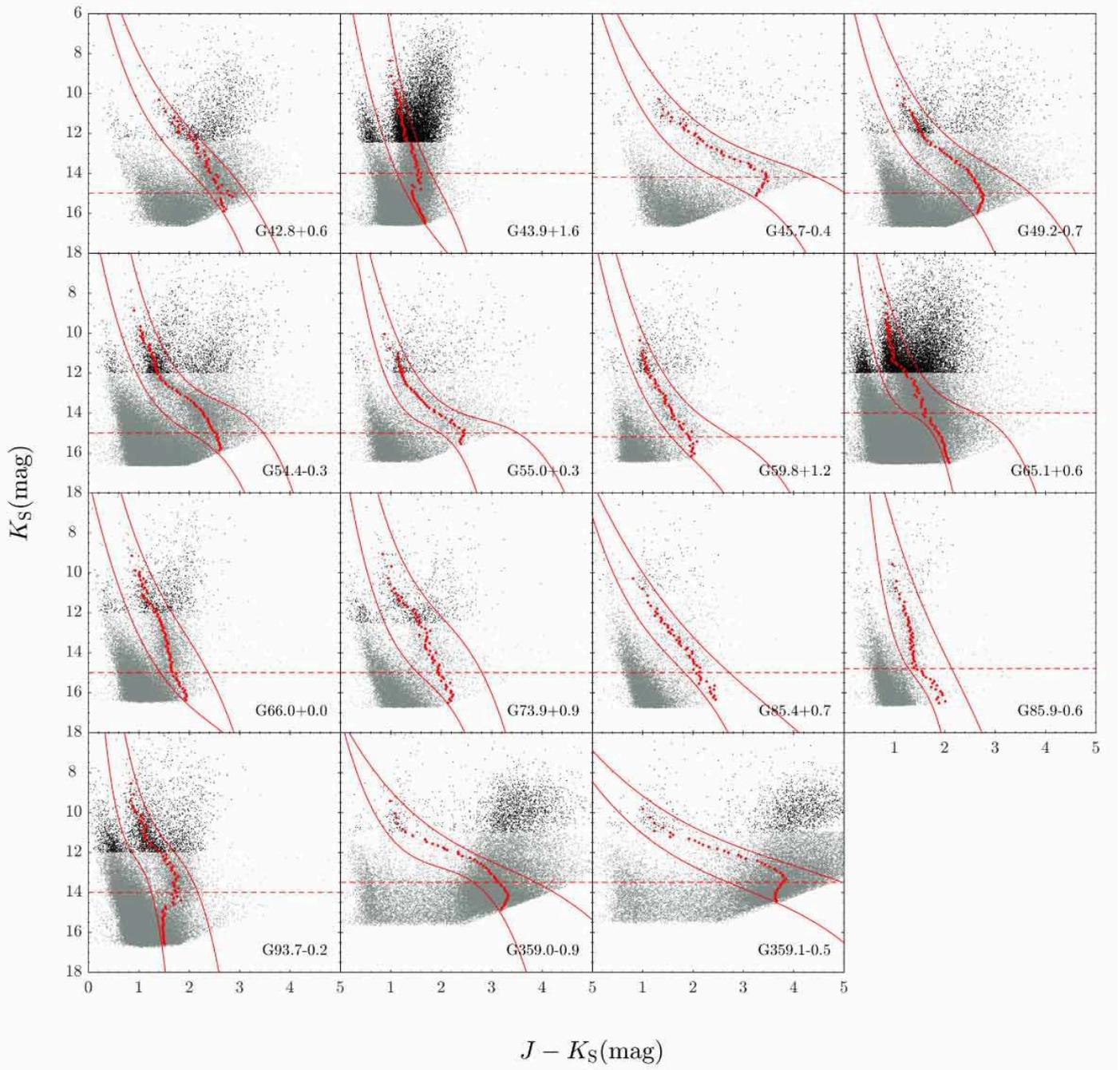}
      \vspace{-1.5in}
   \caption{Same CMDs as Figure \ref{FigUKICMD-1} for SNRs in UKIDSS.}
   \label{FigUKICMD-2}
   \end{figure*}

\begin{figure*}
   \centering
   \includegraphics[width=\hsize]{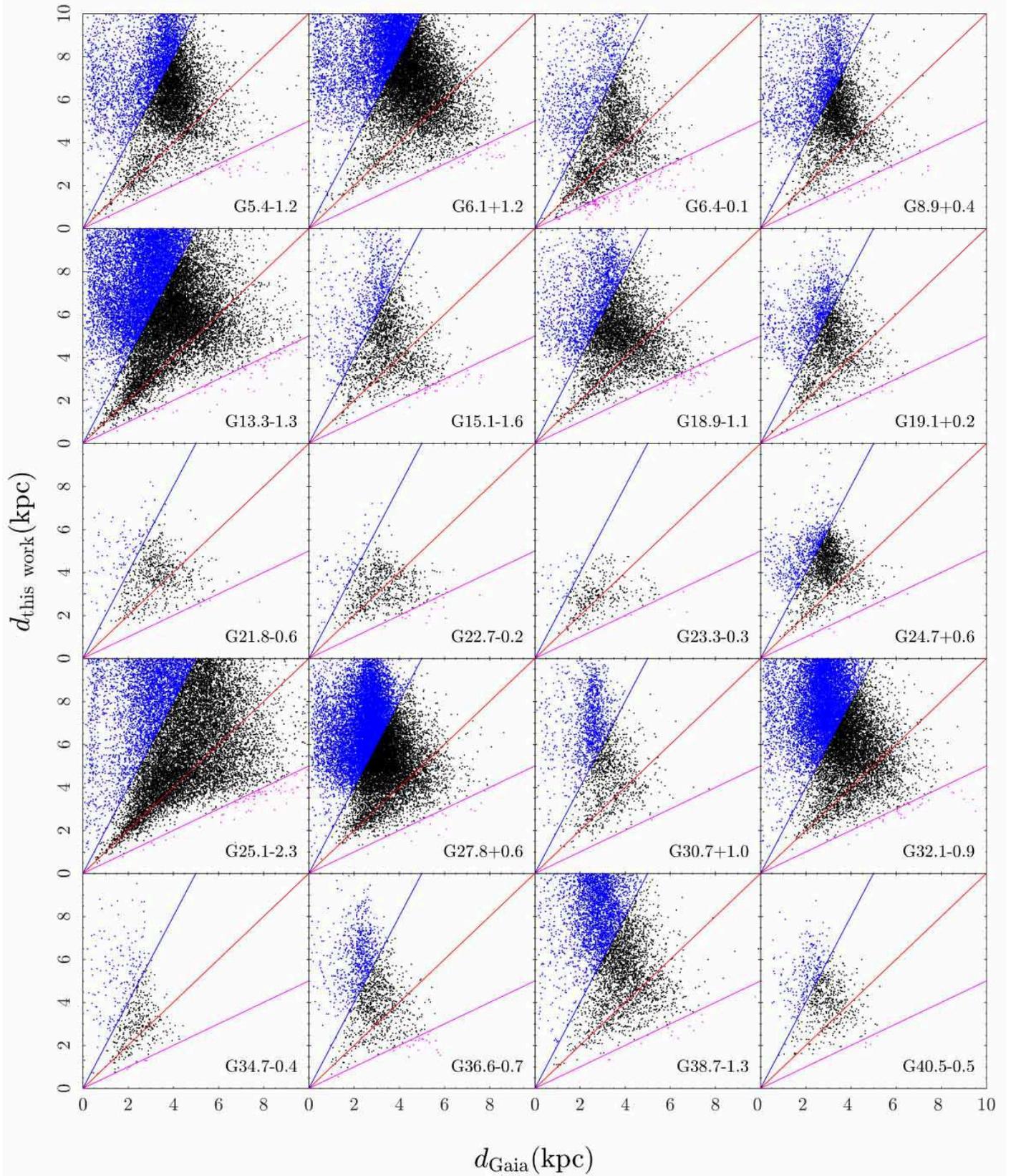}
      \caption{Distance comparison diagrams for SNRs in UKIDSS.
      $d_{\rm this}$ $_{\rm work}$ is calculated under the assumption that stars within the RC strip are RCs.
      $d_{\rm Gaia}$ is collected from {\it Gaia} parallaxes.
      The $y=2x$ (blue) and $y=x/2$ (magenta) lines are used to remove dwarfs and giants, respectively.
      Black points denote RCs, while blue points and magenta points denote dwarfs and giants, respectively.}
   \label{FigUKIDD-1}
   \end{figure*}

\begin{figure*}
   \centering
   \includegraphics[width=\hsize]{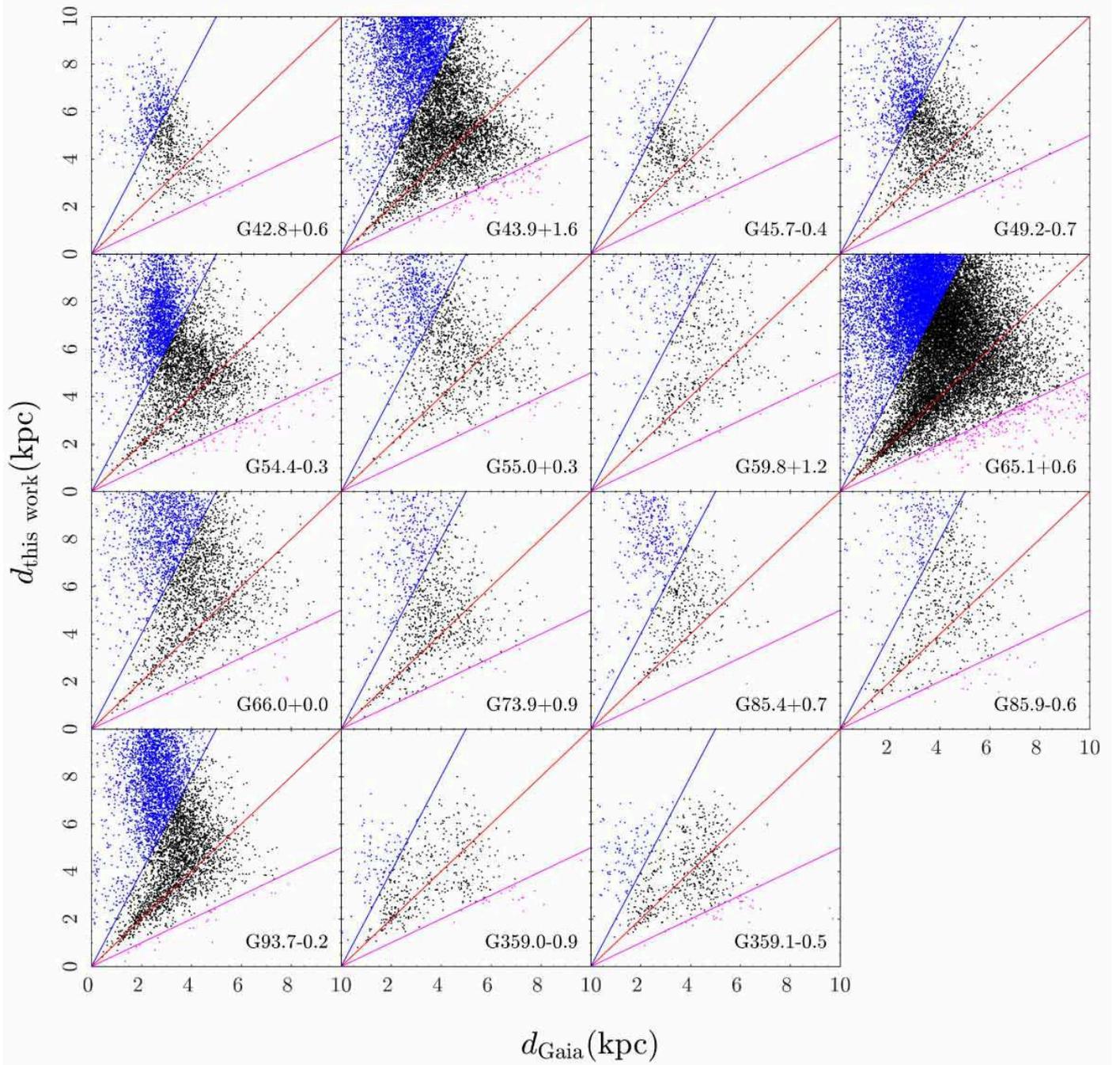}
   \vspace{-1.5in}
      \caption{Same distance comparison diagrams as Figure \ref{FigUKIDD-1} for SNRs in UKIDSS.}
   \label{FigUKIDD-2}
   \end{figure*}

\begin{figure*}[htbp]
   \centering
   \includegraphics[width=\hsize]{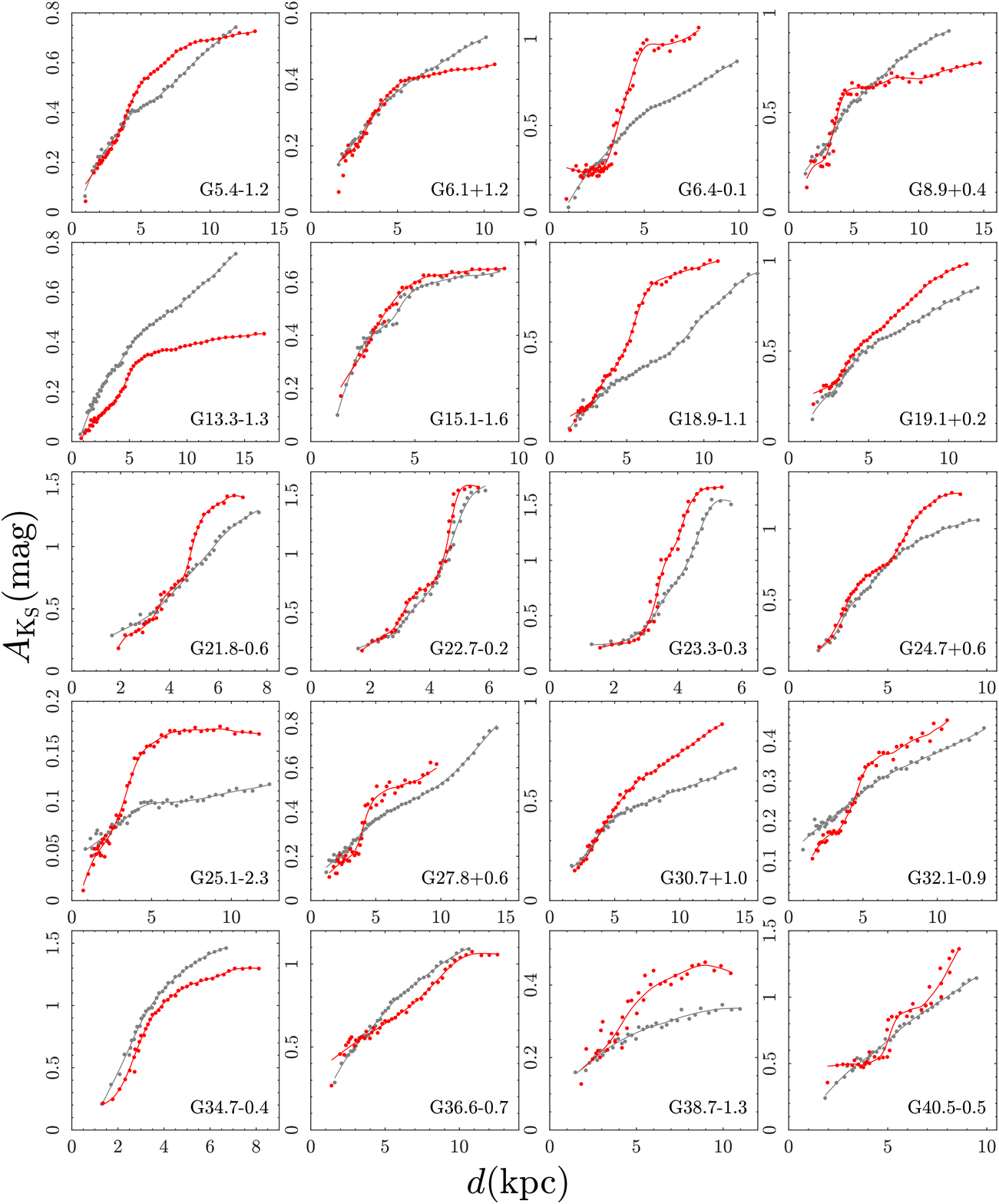}
      \caption{Distance--extinction diagrams for SNRs in UKIDSS.
       Red and grey dots denote the RC ridge of the SNR region and the surrounding region, respectively.
       The lines are the spline interpolation function curves.}
   \label{FigUKIAk-1}
   \end{figure*}

\begin{figure*}[htbp]
   \centering
   \includegraphics[width=\hsize]{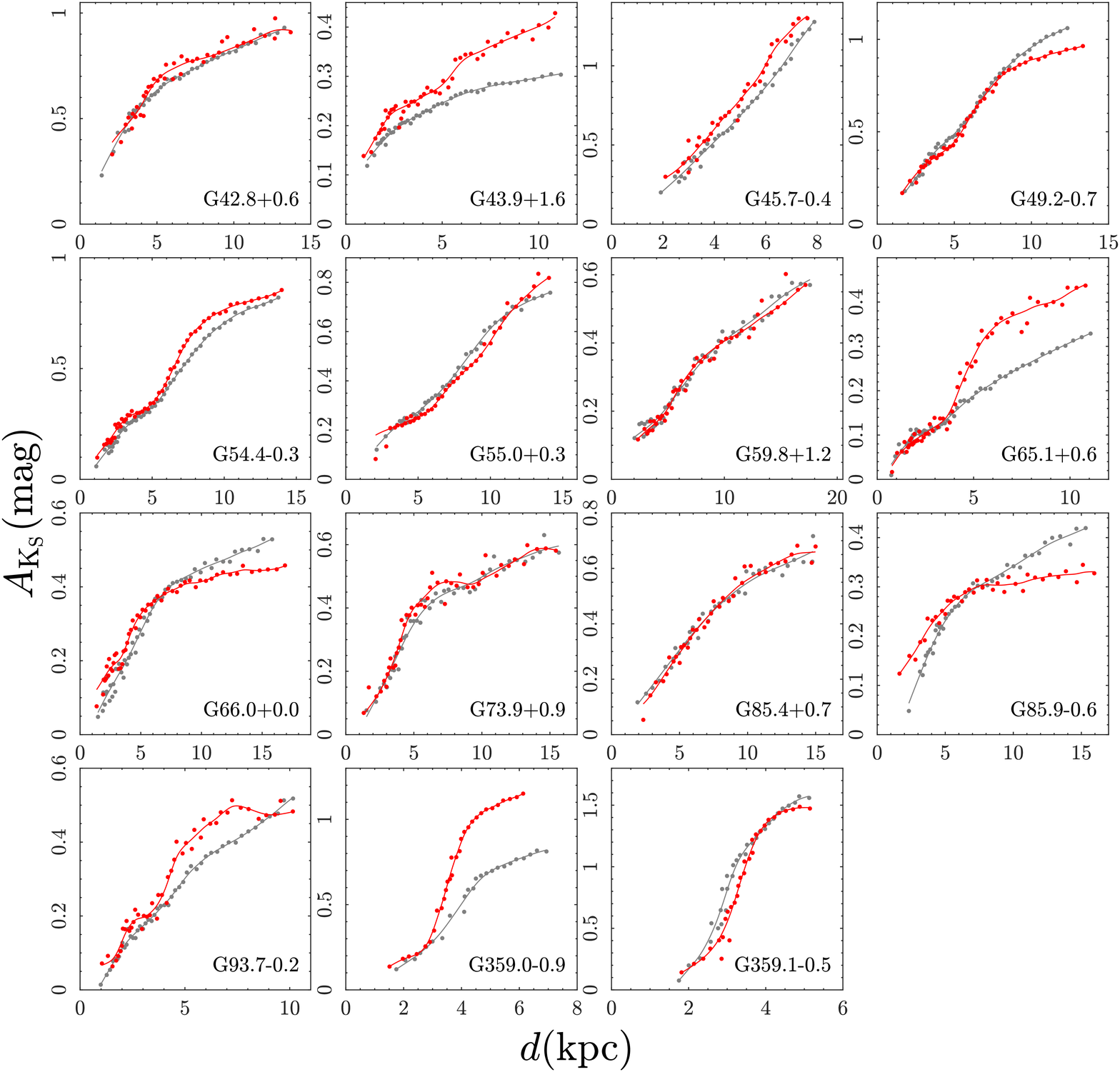}
    \vspace{-1.5in}
       \caption{Same distance--extinction diagrams as Figure \ref{FigUKIAk-1} for SNRs in UKIDSS.}
   \label{FigUKIAk-2}
   \end{figure*}

\begin{figure*}[htbp]
   \centering
   \includegraphics[width=\hsize]{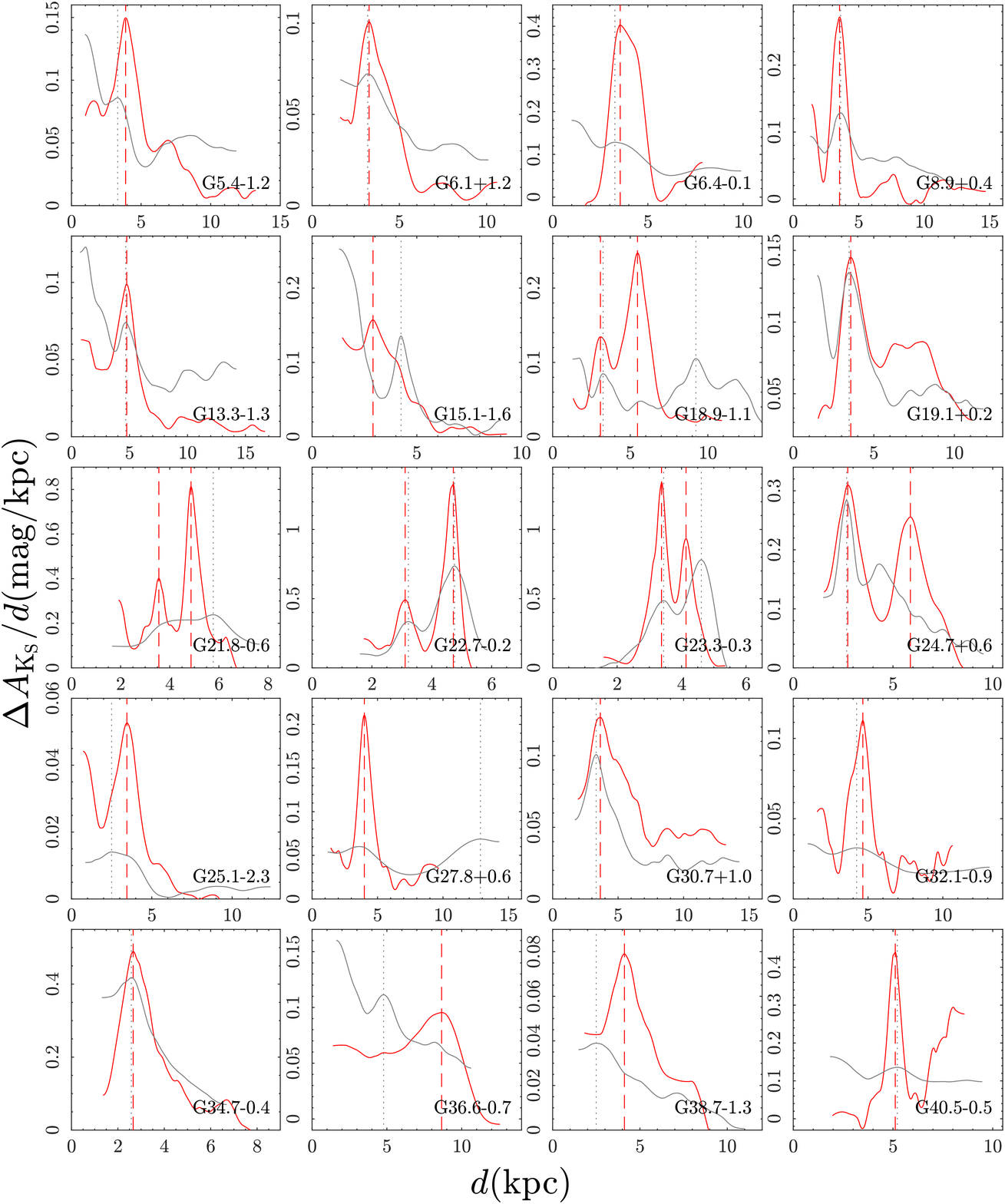}
      \caption{Distributions of differential extinction per distance $\Delta A_\Ks/d$ with distance $d$ for SNRs (red solid lines) and corresponding surrounding areas (grey solid lines) in UKIDSS.
      Red dash lines and grey dotted lines mark the locations of significant gradients.}
   \label{FigUKIdAk-1}
   \end{figure*}

\begin{figure*}[htbp]
   \centering
   \includegraphics[width=\hsize]{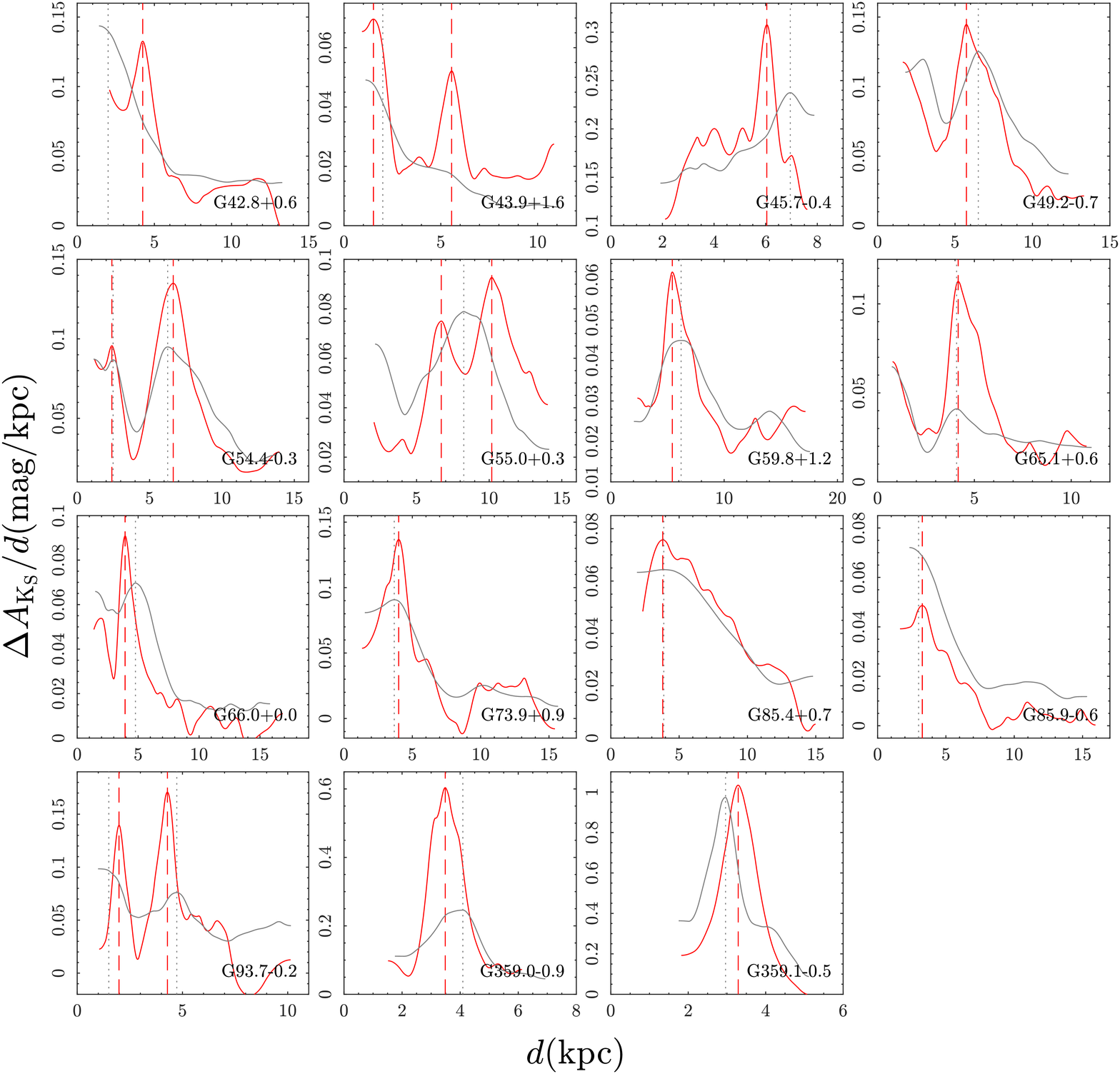}
    \vspace{-1.5in}
      \caption{Same differential extinction per distance--distance diagrams as Figure \ref{FigUKIdAk-1} for SNRs in UKIDSS.}
   \label{FigUKIdAk-2}
   \end{figure*}

\begin{figure*}
   \centering
   \includegraphics[width=\hsize]{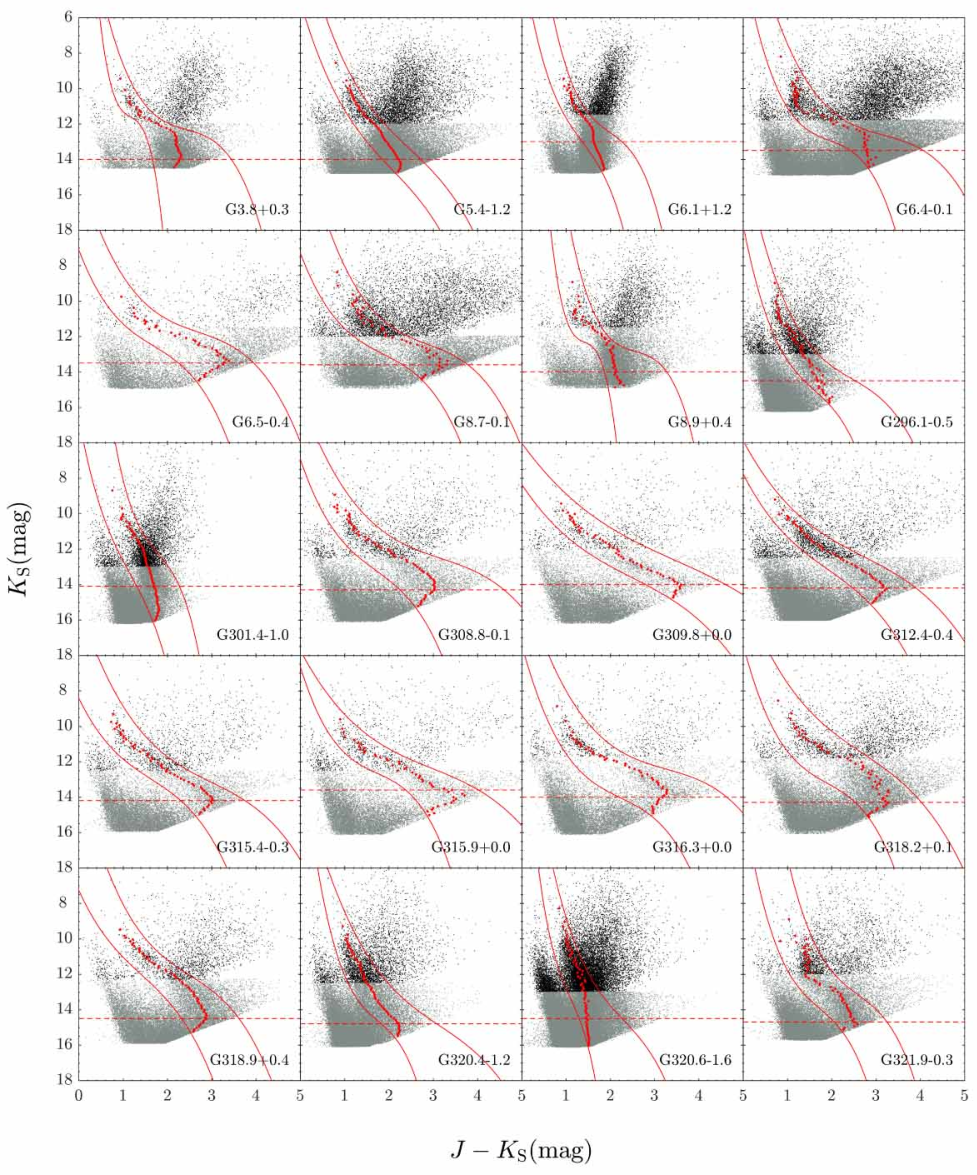}
      \caption{Same CMDs as Figure \ref{FigUKICMD-1}, but for SNRs in VVV.}
   \label{FigVVVCMD-1}
   \end{figure*}

\begin{figure*}[htbp]
   \centering
   \includegraphics[width=\hsize]{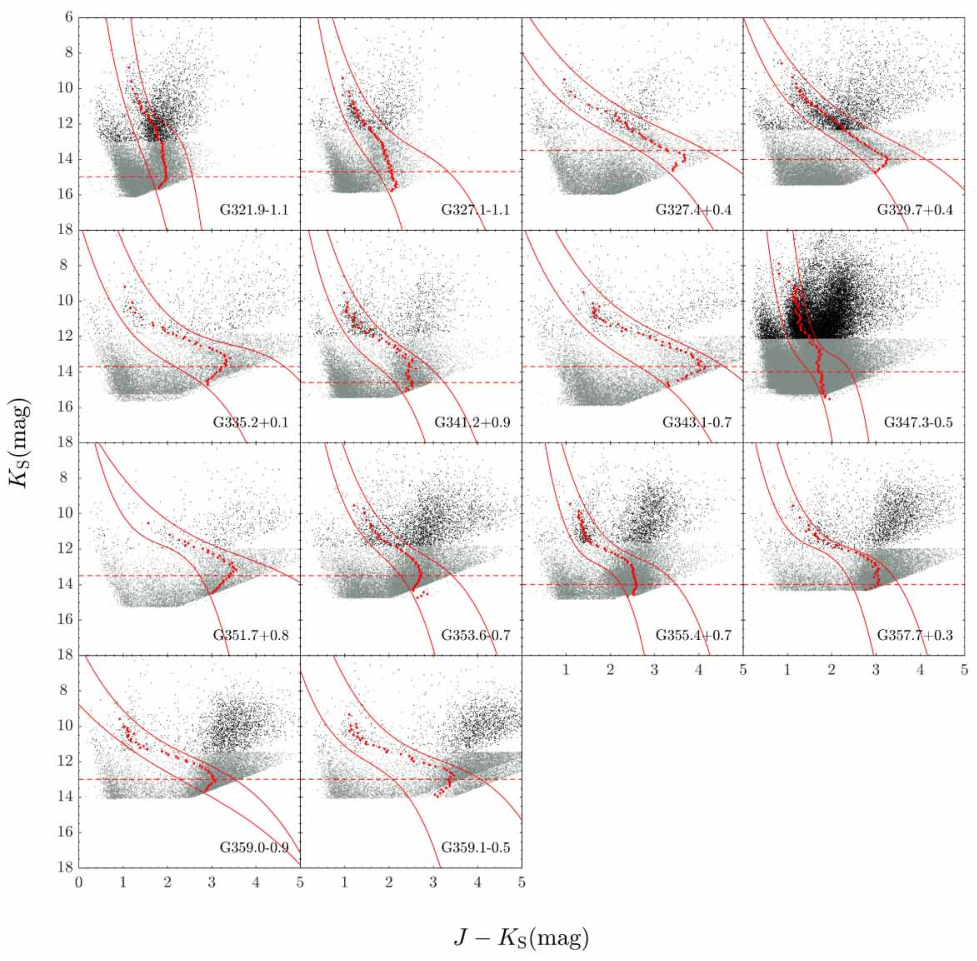}
    \vspace{-1.5in}
      \caption{Same CMDs as Figure \ref{FigUKICMD-1}, but for SNRs in VVV.}
   \label{FigVVVCMD-2}
   \end{figure*}

\begin{figure*}
   \centering
   \includegraphics[width=\hsize]{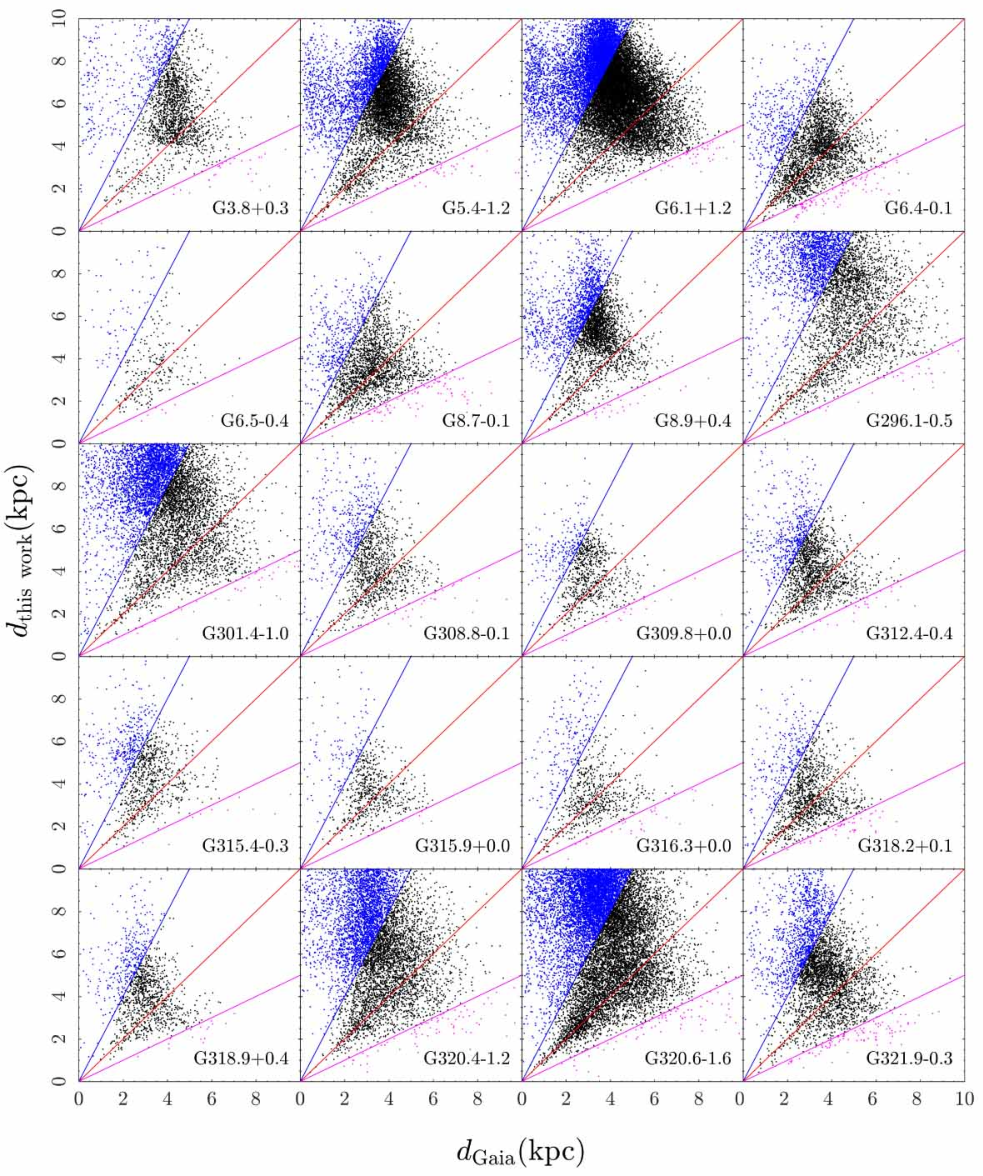}
      \caption{Same distance comparison diagrams as Figure \ref{FigUKIDD-1}, but for SNRs in VVV.}
   \label{FigVVVDD-1}
   \end{figure*}

\begin{figure*}
   \centering
   \includegraphics[width=\hsize]{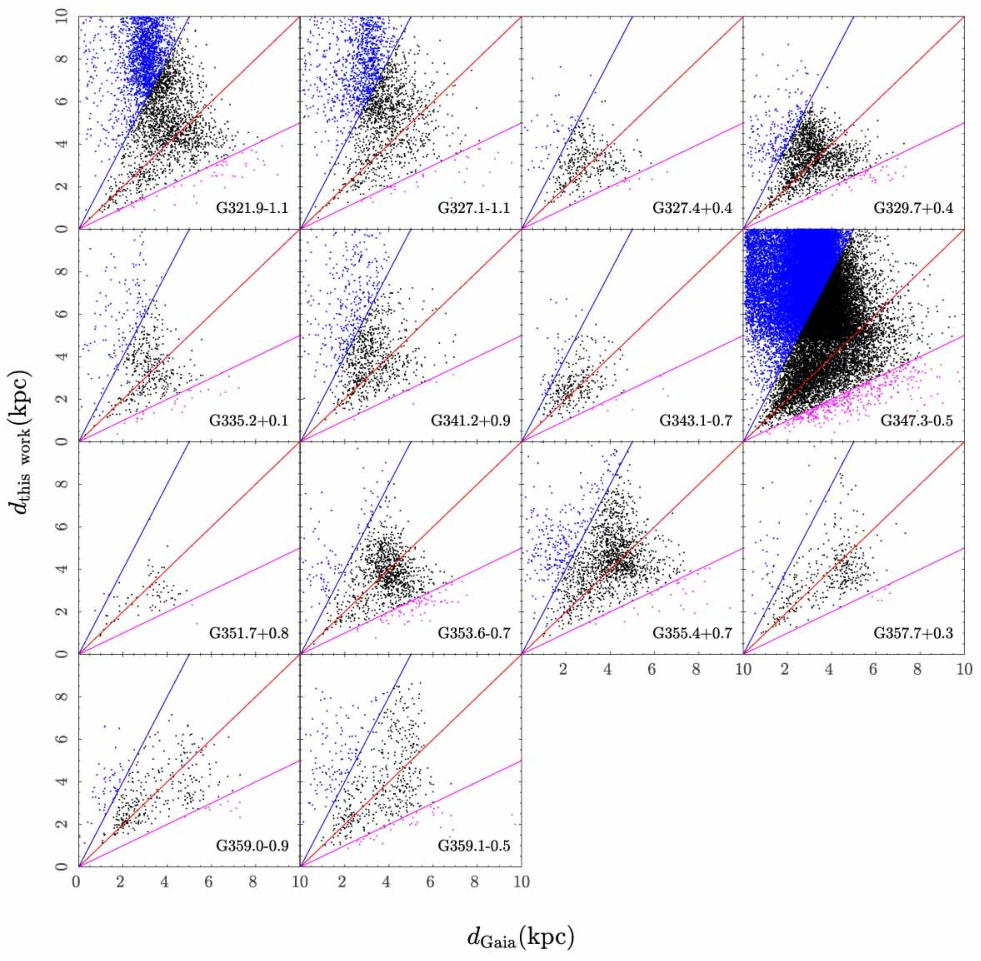}
    \vspace{-1.5in}
      \caption{Same distance comparison diagrams as Figure \ref{FigUKIDD-1}, but for SNRs in VVV.}
   \label{FigVVVDD-2}
   \end{figure*}

\begin{figure*}[htbp]
   \centering
   \includegraphics[width=\hsize]{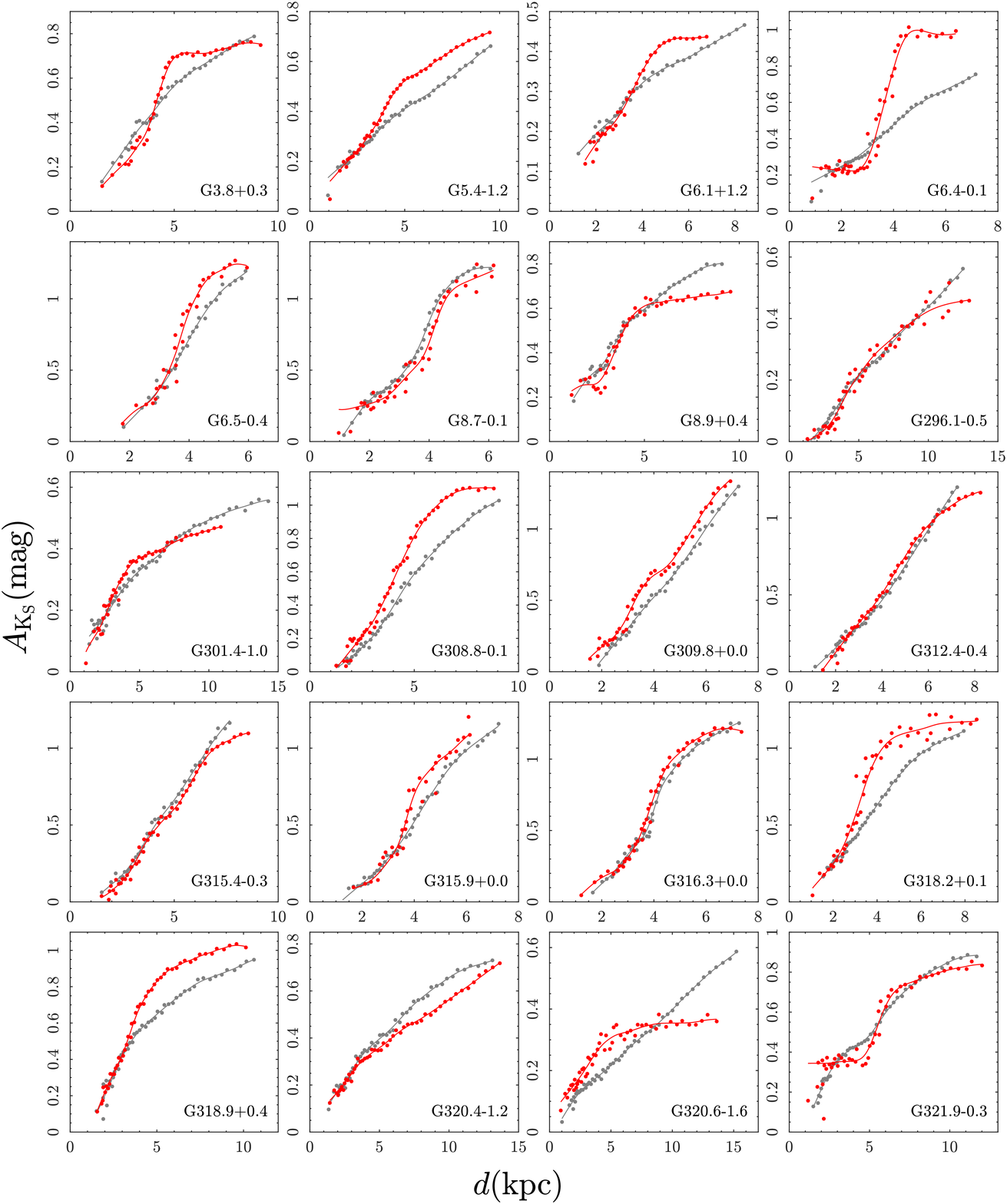}
         \caption{Same distance--extinction diagrams as Figure \ref{FigUKIAk-1}, but for SNRs in VVV.}
   \label{FigVVVAk-1}
   \end{figure*}

\begin{figure*}[htbp]
   \centering
   \includegraphics[width=\hsize]{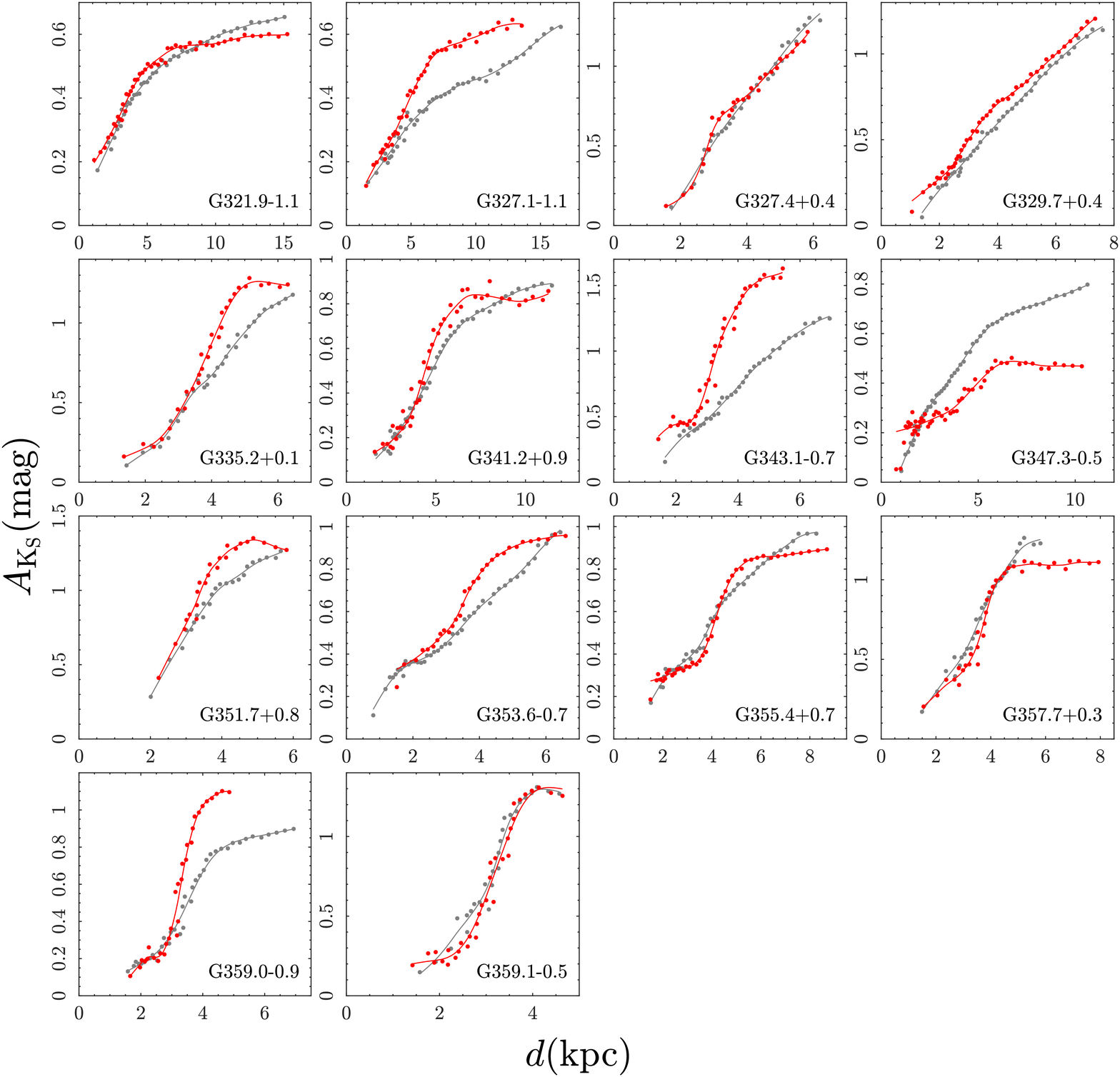}
    \vspace{-1.5in}
      \caption{Same distance--extinction diagrams as Figure \ref{FigUKIAk-1}, but for SNRs in VVV.}
   \label{FigVVVAk-2}
   \end{figure*}

\begin{figure*}[htbp]
   \centering
   \includegraphics[width=\hsize]{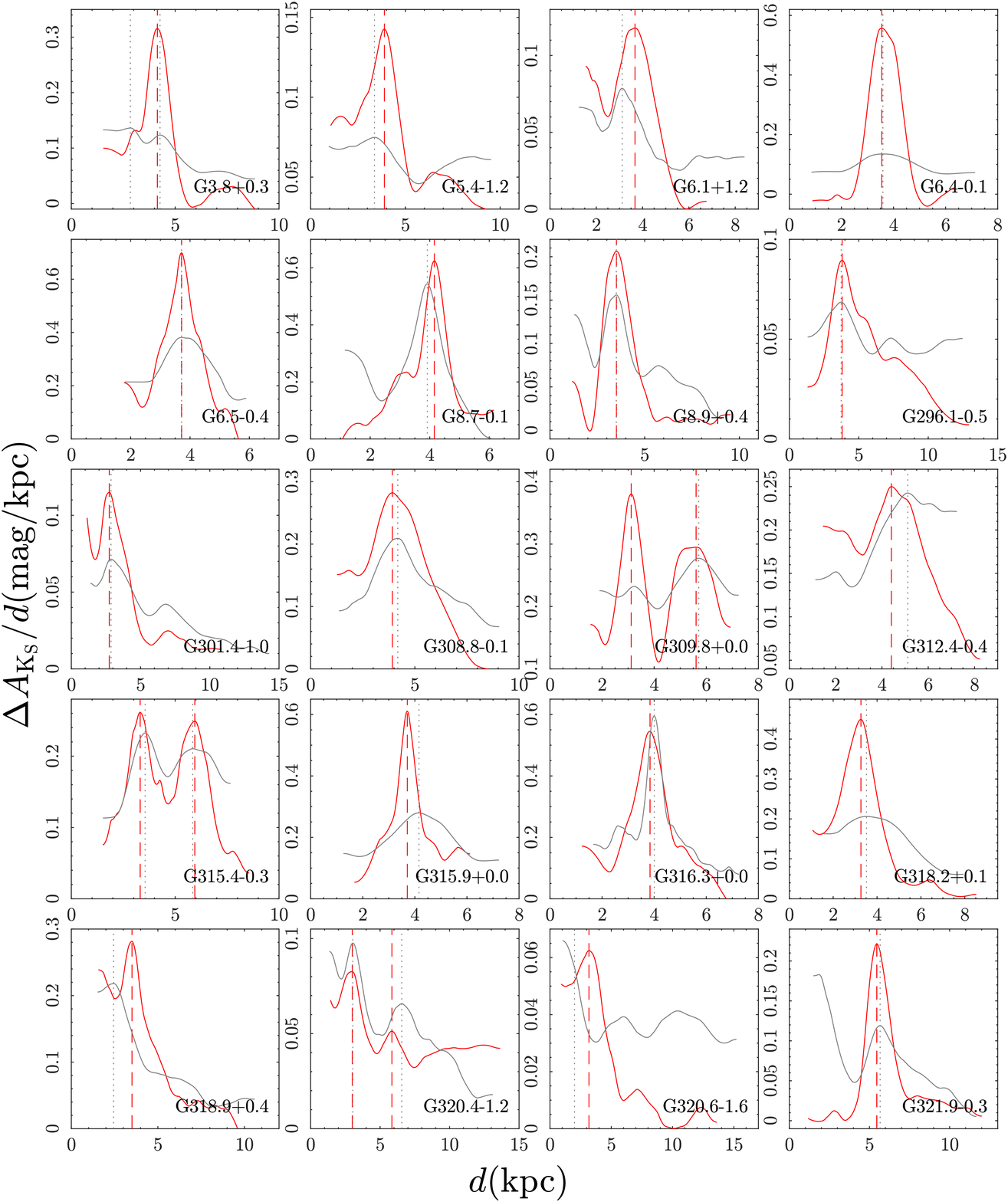}
      \caption{Same differential extinction per distance--distance diagrams as Figure \ref{FigUKIdAk-1}, but for SNRs in VVV.}
   \label{FigVVVdAk-1}
   \end{figure*}

\begin{figure*}[htbp]
   \centering
   \includegraphics[width=\hsize]{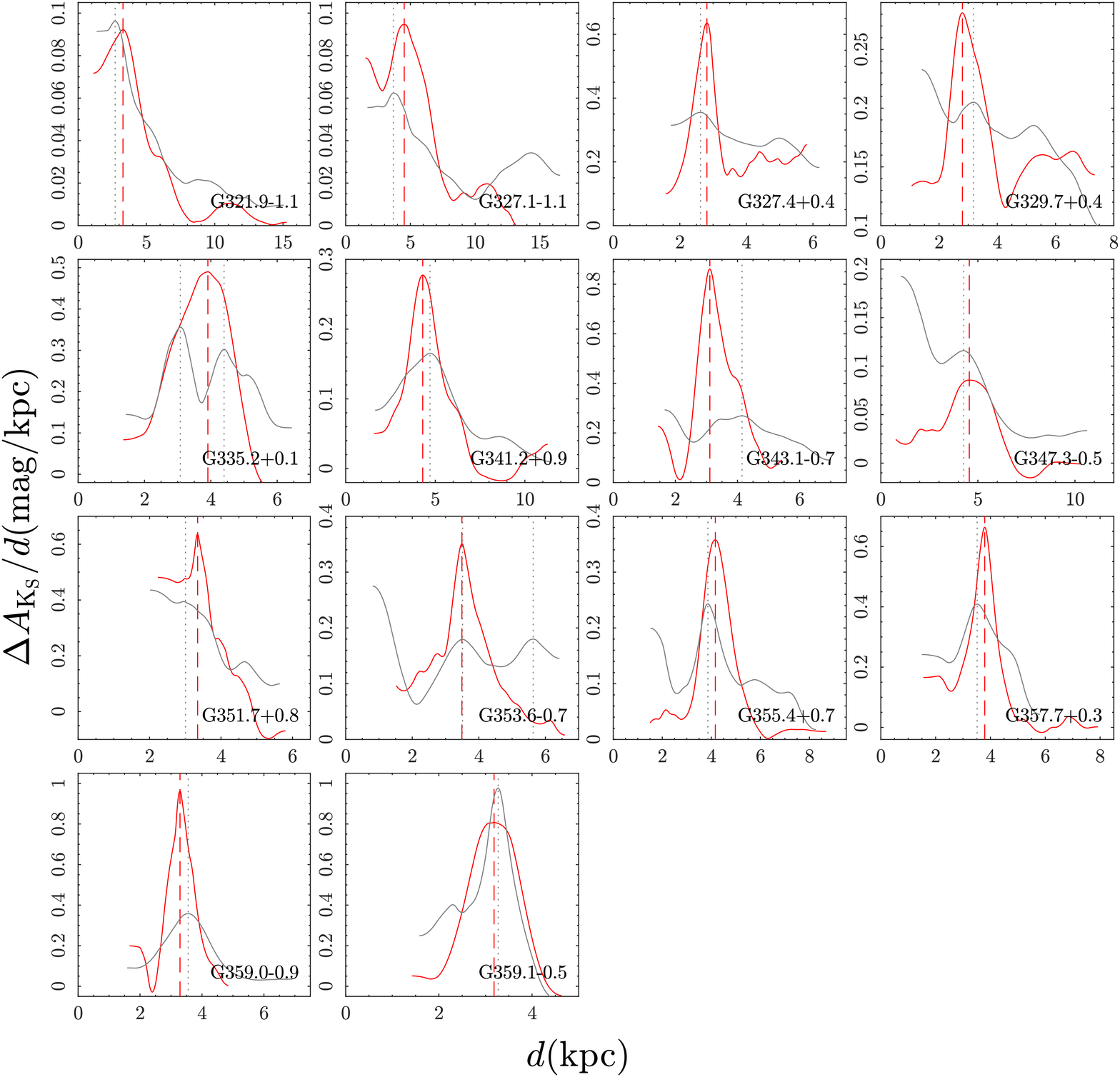}
    \vspace{-1.5in}
      \caption{Same differential extinction per distance--distance diagrams as Figure \ref{FigUKIdAk-1}, but for SNRs in VVV.}
   \label{FigVVVdAk-2}
   \end{figure*}

Figures \ref{FigUKICMD-1}, \ref{FigUKICMD-2}, \ref{FigVVVCMD-1}, and \ref{FigVVVCMD-2}
present the color-magnitude diagrams (CMDs) for all the SNRs.
Figures \ref{FigUKIDD-1}, \ref{FigUKIDD-2}, \ref{FigVVVDD-1}, and \ref{FigVVVDD-2}
are the distance comparison diagrams for all the SNRs.
Figures \ref{FigUKIAk-1}, \ref{FigUKIAk-2}, \ref{FigVVVAk-1}, and \ref{FigVVVAk-2}.
show the distance-extinction diagrams for all the SNRs and the surrounding regions.
Figures \ref{FigUKIdAk-1}, \ref{FigUKIdAk-2}, \ref{FigVVVdAk-1}, and \ref{FigVVVdAk-2}.
display the distributions of differential extinction per distance with distance for all the SNRs and the surrounding regions.

\clearpage

\clearpage

\clearpage

\clearpage

\clearpage

\clearpage

\clearpage

\clearpage

\clearpage

\clearpage

\clearpage

\clearpage

\clearpage

\clearpage

\clearpage

\end{appendix}

\end{document}